\documentclass[preprint,12pt]{elsarticle}
\usepackage{graphicx}
\usepackage{amssymb}
\usepackage{latexsym}
\usepackage{xcolor}

\begin{document}

\begin{frontmatter}
%
%

\title{Numerical method for hydrodynamic modulation equations describing
       Bloch oscillations in semiconductor superlattices}

\author{M. \'Alvaro}
 \ead{mariano.alvaro@uc3m.es}
\author{M. Carretero}
 \ead{manuel.carretero@uc3m.es}
\author{L. L. Bonilla\corref{cor1}}
 \ead{bonilla@ing.uc3m.es}
 \cortext[cor1]{Corresponding author.}
 \address{Gregorio Mill\'an Institute for Fluid Dynamics, Nanoscience and Industrial Mathematics, Universidad Carlos III de Madrid,
Avenida de la Universidad 30, 28911 Legan\'es, Spain} \ead[url]{http://scala.uc3m.es}

\begin{abstract}
We present a finite difference method to solve a new type of nonlocal hydrodynamic equations that arise in the theory of spatially
inhomogeneous Bloch oscillations in semiconductor superlattices. The hydrodynamic equations describe the evolution of the electron density,
electric field and the complex amplitude of the Bloch oscillations for the electron current density and the mean energy density. These
equations contain averages over the Bloch phase which are integrals of the unknown electric field and are derived by singular perturbation
methods. Among the solutions of the hydrodynamic equations, at a 70 K lattice temperature, there are spatially inhomogeneous Bloch
oscillations coexisting with moving electric field  domains and Gunn-type oscillations of the current. At higher temperature (300 K) only
Bloch oscillations remain. These novel solutions are found for restitution coefficients in a narrow interval below their critical values
and disappear for larger values.
We use an efficient numerical method based on an implicit second-order finite difference scheme for both the electric field equation
(of drift-diffusion type) and the parabolic equation for the complex amplitude. Double integrals appearing in the nonlocal hydrodynamic equations
are calculated by means of expansions in modified Bessel functions. We use numerical simulations to ascertain the convergence of the method.
If the complex amplitude equation is solved using a first order scheme for restitution coefficients near their critical values, a spurious convection
arises that annihilates the complex amplitude in the part of the superlattice that is closer to the cathode. This numerical artifact disappears
if the space step is appropriately reduced or we use the second-order numerical scheme.
\end{abstract}

\end{frontmatter}


\setcounter{equation}{0}
\section{Introduction}
An immediate consequence of the Bloch theorem is that the position of an electron inside an energy band of a crystal oscillates coherently under an applied constant electric field with a frequency proportional to the field \cite{zener}. To observe these so-called Bloch oscillations (BOs), their period must be much shorter than the scattering time for, otherwise, scattering eliminates them. The required electric field to observe Bloch oscillations is too large for a natural crystal, but it becomes reasonable if artificial crystals with larger spatial periods are created. An artificial crystal can be formed by growing many identical periods comprising a number of layers of two different materials. The resulting superlattice (SL) was suggested by Esaki and Tsu in 1970 as a possible realization of Bloch oscillations \cite{esaki}. Damped Bloch oscillations were first observed in 1992 in semiconductor SLs whose initial state was prepared optically \cite{fel92}. In recent years, BOs have been observed in other artificial crystals such as atoms placed in the potential minima of a laser-induced optical standing wave \cite{dah96,wil96}, photons in a periodic array of waveguides \cite{per99,mor99} and Bose-Einstein condensates in optical lattices \cite{gre01} among other systems \cite{leo}.

In SLs made out of doped semiconductors, scattering usually destroys BOs, but we have shown recently that BOs can persist even in the hydrodynamic regime for a SL with long scattering times \cite{BACeps11,BAC11}. To do so, we consider a Boltzmann-Poisson description of a SL with a single occupied electron miniband and a dissipative Bhatnagar-Gross-Krook (BGK) collision model \cite{BC08}. In semiconductor SLs, collisions are inelastic: they conserve charge but dissipate energy and momentum. Then the BGK inelastic-collision model contains two restitution coefficients that regulate the fractions of energy and momentum lost in collisions \cite{BC08}. Previously, an inelastic BGK model was used to describe granular fluids, in which energy (but not momentum) is lost during collisions \cite{BMD}. Boltzmann-BGK mass- and energy-conserving kinetic equations with different local equilibrium distributions describe a particle in an external potential, and have been used to derive energy-transport equations and to prove H theorems; see \cite{aok07} and references cited therein. Quantum hydrodynamic and quantum energy-transport models have been derived from quantum BGK kinetic equations that conserve mass, momentum and energy; see \cite{degond07} and references cited therein. For SLs, inelastic collisions imply that BOs are damped on a time scale given by the scattering time. Damping modulates the amplitude of the BOs if the scattering time is sufficiently long. In the limit of almost elastic scattering and high electric field, the electron current density and mean energy oscillate at the Bloch frequency, whereas the electron density, the electric field and the envelope of the BOs vary on a slower scale. In this limit, it is possible to derive a set of one-dimensional nonlocal hydrodynamic equations for the electric field and the complex amplitude of the BOs. Their solutions allow to reconstruct the rapidly varying electron current and mean energy densities. The hydrodynamic equations are of a quite novel type: they contain averages over the phase of the BOs which is proportional to the integral over time of the electric field, and therefore the phase is also an unknown to be determined. Appropriate boundary and initial conditions include initiation of the BOs possibly by optical means \cite{fel92}. Numerical solution of the hydrodynamic equations shows stable BOs for appropriate parameter values \cite{BACeps11, BAC11}. For dc voltage biased SLs at sufficiently low lattice temperatures, there are solutions in which the amplitude of the BOs is time-periodic and the electric field profile inside the SL exhibits electric field domains (EFDs) \cite{BAC11}.

In this paper, we present a method to solve numerically the nonlocal hydrodynamic equations describing BOs for a dc voltage biased SL. Although the problem is one-dimensional (1D), it is very computationally intensive, so we need an efficient numerical method to solve it. We solve the hydrodynamics equations  by means of an efficient implicit finite difference numerical scheme, which uses a fixed point iteration process to obtain numerically both the electric field and the BO complex amplitude at each time step. The equation for the field is a nonlocal drift-diffusion equation (containing integrals over the Bloch phase which is an integral of the electric field) that is solved using an implicit numerical scheme involving the inversion of only one tridiagonal matrix per iteration. This equation is coupled to that of the BO amplitude. We use a second order numerical scheme to solve the latter. In the hydrodynamic equations, there appear several Fourier coefficients of the Boltzmann distribution function (which is periodic in the momentum variable with period $2\pi/l$ if $l$ is the spatial period of the SL). These Fourier coefficients are approximated by truncated series of modified Bessel functions for computational efficiency. With appropriate parameter values and boundary conditions, numerical solutions show that initial profiles for the field and the BO amplitude evolve to stable spatially inhomogeneous profiles at room temperature \cite{BAC11}. At low temperature (70 K), we have found that BOs and Gunn-type oscillations \cite{kroemer,BT10} due to EFD dynamics may coexist. Increasing lattice temperature produces large diffusion coefficients as compared to the convective part of the averaged drift-diffusion equation for the electric field. This eliminates the Gunn-type oscillations. At low lattice temperature, the diffusion does not change that much, but convection dominates the average electron current density, thereby facilitating movable EFDs and Gunn-type oscillations \cite{BAC11}. BOs (accompanied or not by Gunn-type oscillations in their amplitude) disappear as the collisions in the kinetic equation become more inelastic and the BO amplitude becomes zero everywhere. If the amplitude of the BOs is set to zero, the drift-diffusion equation for the electric field is similar to those obtained with a local equilibrium distribution that depends only on the electron density \cite{BEP03}. Solutions of this drift-diffusion equation include  Gunn-type self-oscillations due to EFD dynamics. Direct solution of the Boltzmann-Poisson system studied in \cite{BEP03} confirms this \cite{elena}.

The rest of the paper is as follows. In Section \ref{sec:2}, we describe the Boltzmann-BGK-Poisson system and the nondimensional hydrodynamic equations derived from it. In Section \ref{sec:3}, we explain the numerical method for solving the  hydrodynamic equations as well as the numerical results. The analysis of the convergence of the numerical method is based on numerical simulations and it is presented in section \ref{sec:4}. Section \ref{sec:5} contains our conclusions. In \ref{app} we include some technical details including the series of modified Bessel functions used to approximate some integrals.

\section{Model equations} \label{sec:2}
For a semiconductor superlattice with a single occupied miniband of dispersion relation
\begin{eqnarray}
{\cal E}(k)=\frac{\Delta}{2}(1-\cos kl)\label{eq1}
\end{eqnarray}
($\Delta$ is the SL miniband width and $l$ is the SL spatial period), the distribution function $f(x,k,t)$ of electrons with position in the interval $(x,x+dx)$ and wave vector in $(k,k+dk)$ satisfies the system of equations \cite{BC08,BAC11}:
\begin{eqnarray}
&&\frac{\partial f}{ \partial t} + v(k)\, \frac{\partial f}{ \partial x} +  \frac{eF}{ \hbar} \,\frac{\partial f}{\partial k} = - \nu\,(f - f^{B}),   \label{eq2}\\
&&\varepsilon\, \frac{\partial F}{\partial x} = \frac{e}{ l}\, (n-N_{D}),
\label{eq3}\\
&& n = \frac{ l}{ 2\pi} \int_{-\pi/l}^{\pi/l} f(x,k,t)\, dk,    \label{eq4}\\
&& f^{B}(k;n,u_\alpha,T_\alpha) = n\,\frac{\exp\!\left(\frac{ \hbar k u_{\alpha} - {\cal E}(k)}{k_{B}T_{\alpha}}\right)}{ \frac{ l}{ 2\pi} \int_{-\pi/l}^{\pi/l}\exp\!\left(\frac{ \hbar k u_{\alpha} - {\cal E}(k)}{k_{B}T_{\alpha}}\right)\!dk},  \label{eq5}\\
&& \frac{ l}{ 2\pi} \int_{-\pi/l}^{\pi/l} f^{B}(k;n,u_\alpha,T_\alpha)\, dk= n,    \label{eq6}\\
&& \frac{e}{ 2\pi} \int_{-\pi/l}^{\pi/l} v(k)\, f^{B}\, dk= (1-\alpha_{j}) J_{n},
\label{eq7}\\
&& \frac{l}{ 2\pi n} \int_{-\pi/l}^{\pi/l}  \left[\frac{\Delta}{ 2} -{\cal E}(k)\right]
f^{B}\, dk = \alpha_{e} E_{0} + (1-\alpha_{e}) E, \label{eq8}\\
&& J_n=\frac{e}{ 2\pi} \int_{-\pi/l}^{\pi/l} v(k)\, f\, dk, \label{eq9}\\
&&E=\frac{l}{ 2\pi n} \int_{-\pi/l}^{\pi/l}  \left[\frac{\Delta}{ 2} -{\cal E}(k)\right] f\, dk.
\label{eq10}
\end{eqnarray}
Here $f^B$, $n$, $N_{D}$, $\varepsilon$, $v(k)=\hbar^{-1}d\mathcal{E}/dk$, $k_{B}$, $-e<0$, $u_\alpha$, $T_\alpha$, $J_n$, $E$, $E_0$ and $-F=-\partial W/\partial x$ are the local equilibrium distribution, the 2D electron density, the 2D doping density, the dielectric constant, the group velocity, the Boltzmann constant, the electron charge, the hydrodynamic velocity, the electron temperature, the electron current density, the mean energy density, the lattice mean energy density, and the electric field, respectively. $W$ is the electric potential. The lattice mean energy density is related to the lattice temperature $T_0$ as $E_0=\frac{\Delta}{2} I_1(\frac{\Delta}{2k_BT_0})/I_0(\frac{\Delta}{2k_BT_0})$, where $I_j(x)$ is the modified Bessel function of index $j$ \cite{BAC11}. Note that the 1D distribution functions $f$ and $f^B$ have the same units as the 2D electron density $n$ and that $u_\alpha$ and $T_\alpha$ are functions of $n$, $J_n$ and $E$ obtained by solving (\ref{eq7})-(\ref{eq8}) with $f^B$ given by (\ref{eq5}). The 1D Boltzmann local equilibrium (\ref{eq5}) is an approximation to a more general Fermi-Dirac distribution \cite{BAC11}. $\nu$ is the collision frequency which we take as a constant for the sake of simplicity. $\alpha_e$ and $\alpha_j$ are constant restitution coefficients that indicate the fraction of energy and momentum dissipated in inelastic collisions (with phonons, for example). The distribution function is periodic in $k$ with period $2\pi/l$. 

Ktitorov, Simin and Sindalovskii (KSS) \cite{KSS} proposed in 1972 an equation similar to (\ref{eq2}) except that $f^B$ was replaced by the Boltzmann equilibrium distribution at the lattice temperature $T_0$ and an additional term $Q_p=-\nu_p[f(x,k,t)-f(x.-k,t)]/2$ (representing 1D impurity collisions) was added to the RHS. Later Ignatov and Shashkin \cite{ISh87} proposed a local equilibrium distribution function similar to (\ref{eq5}) with $u_\alpha=0$ and $T_\alpha=T_0$. Such a kinetic model was numerically solved by Cebri\'an et al \cite{elena} using a particle method, and Bonilla et al \cite{BEP03} derived from it a generalized drift-diffusion equation for the electric field exhibiting Gunn-like oscillations of the current due to EFD dynamics. However the KSS model with the Ignatov-Shashkin modification cannot sustain stable Bloch oscillations \cite{BAC11}. To see the relation with BOs, we multiply (\ref{eq2}) by 1, $v(k)$ or $[\Delta/2-\mathcal{E}(k)]$ and integrate over $k$, thereby obtaining the following moment equations for $n$, $J_n$ and $E$:
\begin{eqnarray}
&& \frac{e}{l}\,\frac{\partial n}{\partial t} + \frac{\partial J_{n}}{\partial
x} = 0,  \label{eq11}\\
&& \frac{\partial J_{n}}{ \partial t} + \frac{e\Delta^2 l}{ 8\hbar^2}\,
\frac{\partial}{\partial x}(n-\mbox{Re}\, f_{2}) - \frac{e^2l\, nEF}{\hbar^2} = -
\nu\alpha_{j} J_{n},    \label{eq12}\\
&& \frac{\partial E}{\partial t} - \frac{l E}{en}\,\frac{\partial J_{n}}{
\partial x} - \frac{\Delta^2l}{8\hbar n}\,\frac{\partial}{\partial x} \mbox{Im}
f_{2} + \frac{F\, J_{n}l}{n} = - \nu\alpha_e (E-E_0).   \label{eq13}
\end{eqnarray}
Here we have used (\ref{eq1}) and the Fourier coefficients $f_{j}$ of the periodic distribution function:
\begin{equation}
f(x,k,t) = \sum_{j=-\infty}^\infty f_{j}(x,t)\, e^{ijkl}. \label{eq14}
\end{equation}
Note that $J_{n}= - e\Delta$ Im$f_{1}/(2\hbar)$ and $E= \Delta\, \mbox{Re} f_{1}/(2n)$. (\ref{eq11}) is the charge continuity equation. For elastic collisions, $\alpha_e=\alpha_j= 0$ and space-independent $n$, $J_n$ and $E$, we obtain $\partial n/\partial t=0$ (thus $n$ is constant), and (\ref{eq12}) and (\ref{eq13}) become
\begin{eqnarray}
&& \frac{\partial J_{n}}{ \partial t} - \frac{e^2l\, nEF}{\hbar^2} = 0,    \label{eq15}\\
&& \frac{\partial E}{\partial t} + \frac{F\, J_{n}l}{n} = 0.   \label{eq16}
\end{eqnarray}
Since $n$ is constant, $J_n$ and $E$ are time periodic and oscillate with the Bloch frequency $\omega_B=eFl/\hbar$, proportional to the electric field. In the general case of space dependent moments, (\ref{eq11})-(\ref{eq13}) are not a closed system of equations because they depend on the second harmonic of the distribution function $f_2$. In an appropriate limit, the terms on the RHS of (\ref{eq12}) and (\ref{eq13}) modulate the BOs, so that $n$, $F$ and the amplitude of the BOs evolve on a slower time scale. Based on these ideas, we have derived in \cite{BAC11} a system of slowly-varying nonlocal hydrodynamic equations for these magnitudes using singular perturbation methods.

\begin{table}[ht]
\begin{center}\begin{tabular}{ccccccccc}
 \hline
$[f]$, $[n]$, $[A]$ & $[F]$ &$[{\cal E}]$, $[E]$ &$[v(k)]$&$[J_{n}]$& $[x]$ & $[k]$ & $[t]$ & $\delta$\\
$N_{D}$ & $\frac{\hbar\nu}{el}$ & $\frac{\Delta}{2}$ & $\frac{l\Delta}{2
\hbar}$ & $\frac{eN_{D}\Delta}{2\hbar}$& $\frac{\varepsilon\hbar\nu}{
e^2N_{D}}$ & $\frac{1}{l}$ & $\frac{2\varepsilon\hbar^2\nu}{e^2N_{D}l
\Delta}$ & $\frac{e^2N_D l\Delta}{2\varepsilon\hbar^2\nu^2}$\\
$10^{10}$cm$^{-2}$ & kV/cm & meV & $10^4$m/s
& $10^4$A/cm$^2$ & nm & 1/nm & ps & -- \\
$4.048$& 130 & 8 & 6.15 & 7.88 & 116 & 0.2 & 1.88& 0.0053\\
 \hline
\end{tabular}
\end{center}
\caption{Hyperbolic scaling and nondimensionalization with $\nu=10^{14}$ Hz.}
\label{t1}
\end{table}

\subsection{Hydrodynamic equations}
Written in the nondimensional units of table \ref{t1}, the hydrodynamic equations are \cite{BAC11}
\begin{eqnarray}
&&\frac{\partial F}{\partial t} +\frac{\delta}{F^2+\delta^2\gamma_{j}\gamma_e} \left[ \gamma_{e}E_{0}nF+\frac{F}{2}\, \frac{\partial}{\partial x}\mbox{Im}\,\frac{f^{B(0)}_{2,0}}{1+2iF}\right.\nonumber\\
&& \quad\left. - \frac{\delta\gamma_{e}}{2}\,\frac{\partial}{\partial x}\left(n-\mbox{Re}\,\frac{f^{B(0)}_{2,0}}{1+2iF}\right) - F
\mbox{Re}\, h_{S}+\delta
\gamma_{e}\mbox{Im}\, h_{S}\right] =\langle J\rangle_\theta(t), \label{ddeF}\\
&&\frac{\partial A}{\partial t} =-\frac{\gamma_{e}+\gamma_{j}}{2} \,A + \frac{1}{2i}\frac{\partial}{\partial
x}\!\left(\frac{f^{B(0)}_{2,-1} + \delta\, r^{(1)}_{2,-1}}{1+iF}\right)\!,  \label{eqA}\\
&&J(t,\theta) = \langle J\rangle_\theta-\frac{1}{L}\int_0^L\mbox{Im }(A e^{-i\theta})\, dx, \label{eqJtot}\\
&&n=1+\frac{\partial F}{\partial x},  \label{poiss}
\end{eqnarray}
where
\begin{eqnarray}
&&r^{(1)}_{2,-1} = f^{B(1)}_{2,-1}\! -\!\left(\mathcal{A}^{(0)} \frac{\partial}{\partial A}+(\langle
J\rangle_\theta-J_{n,Su})\frac{\partial}{\partial F}-\frac{\partial J_{n,Su}}{\partial x}\frac{\partial}{\partial
n}\right)\!\frac{f^{B(0)}_{2,-1}}{1+iF}\nonumber\\
&&\quad\quad -\frac{1}{2i}\frac{\partial}{\partial x}\!\left(A -\frac{f^{B(0)}_{3,-1}}{1+2iF}\right)\!, \label{r2m1}\\
&&\mathcal{A}^{(0)} = -\frac{1}{2}(\gamma_{e}+\gamma_{j})\, A + \frac{1}{2i}\,\frac{\partial }{\partial
x}\frac{f^{B(0)}_{2,-1}}{1+iF} ,  \nonumber\\
&&J_{n,Su}= {\delta\gamma_e nE_0F \over \delta^2 \gamma_j\gamma_{e}+F^2}, \nonumber\\
&&h_S =  - \frac{\delta\gamma_{e}E_{0}(\delta\gamma_{j}-iF)}{\delta^2\gamma_{e}\gamma_{j}+ F^2}\,  \frac{\partial J_{n,Su}}{\partial x} +
(\langle J\rangle_\theta-J_{n,Su})\,\frac{\partial}{\partial
F}\left(\frac{\delta\gamma_{e}nE_{0}(\delta\gamma_{j}-iF)}{\delta^2\gamma_{e}\gamma_{j}+
F^2}\right), \nonumber\\
&&\theta =\frac{1}{\delta}\,\int_{0}^t F(x,s)\, ds. \label{eqtheta}
\end{eqnarray}
Here $\gamma_{e,j}=\alpha_{e,j}/\delta$ are rescaled restitution coefficients and $f^B_j$ and $f_{j,l}^B$ are the Fourier coefficients,
\begin{eqnarray}
f^{B}_{j} &=& {1 \over 2\pi}\int_{-\pi}^\pi f^{B}\, e^{-ijk}\,dk , \label{fBj}\\
f^{B}_{j,l} &=& {1 \over 2\pi} \int_{-\pi}^\pi  f^{B}_j\, e^{-il\theta}\, d\theta. \label{fBjl}
\end{eqnarray}
of the Boltzmann local equilibrium distribution (\ref{eq5}) which, in nondimensional form, is:
\begin{eqnarray}
&&f^B(n,u,\beta,k) = n\,\frac{\pi\, e^{ uk + \beta\cos k}}{\int_{0}^\pi e^{\beta\cos K}\cosh(uK)\, dK}. \label{eq92}
\end{eqnarray}
The Boltzmann local equilibrium distribution can be expanded in a power series of the small parameter $\delta$:
\begin{eqnarray}
&&f^B = \sum_{m=0}^{\infty} f^{B(m)}\, \delta^m.\label{eq93}
\end{eqnarray}
In (\ref{eq92}), $u$ and $\beta$ are nondimensional multipliers corresponding to $u_\alpha$ and to $1/(k_BT_\alpha)$ in (\ref{eq5}). They are found by solving the nondimensional versions of (\ref{eq7}) and (\ref{eq8}):
\begin{eqnarray}
&& \frac{1}{ 2\pi} \int_{-\pi}^{\pi} f^{B}\sin k\, dk= (1-\delta\gamma_{j}) J_{n}, \label{eq17}\\
&& \frac{1}{ 2\pi n} \int_{-\pi}^{\pi} f^{B}\cos k\, dk = E-\delta\gamma_{e} (E-E_{0}), \label{eq18}
\end{eqnarray}
when $f^B$ is given by (\ref{eq92}). In (\ref{ddeF}) and (\ref{r2m1}), $f^{B(0)}_{1}$ is given by
\begin{eqnarray}
&&f^{B(0)}_{1}=A\, e^{-i\theta}.\label{eqNewton}
\end{eqnarray}
To calculate $f^{B(1)}_{2,-1}$ in (\ref{r2m1}), we need $u^{(1)}$ and $\beta^{(1)}$ given by
\begin{eqnarray}
&& f^{B(1)}_{1}=\gamma_{e}n E_{0}+f_{1,S}-\frac{\gamma_{e}+\gamma_{j}}{2}\, A\,
e^{-i\theta}-\frac{\gamma_{e}-\gamma_{j}}{2}\,\overline{A} e^{i\theta}, \label{equ1beta1}
\end{eqnarray}
where $\overline{A}$ is the complex conjugate of $A$, and
\begin{eqnarray}
f_{1,S} &=& \frac{\delta}{F^2+\delta^2\gamma_{j}\gamma_e}
\left[\gamma_{e}nE_{0} (\delta\gamma_{j}-iF)\right.\nonumber\\
&+& \frac{F+i\delta\gamma_{e}}{2}\, \frac{\partial}{\partial x}\left(n-\mbox{Re}\, \frac{f_{2,0}^{B(0)}}{1+i2F}\right)
- (\delta\gamma_{j}-iF)\mbox{Re}h_{S} \nonumber\\
&+& \left.\frac{\delta\gamma_{j}-iF}{2}\, \mbox{Im}\frac{\partial}{\partial x} \left(\frac{f_{2,0}^{B(0)}}{1+i2F}\right)
-(F+i\delta\gamma_{e})\mbox{Im}h_{S}\right]. \label{cap2.sl8}
\end{eqnarray}

Together with the Poisson equation (\ref{poiss}), (\ref{ddeF}) is a drift-diffusion equation for the electric field $-F$, (\ref{eqA}) gives the time evolution of the BO complex amplitude $A$, and (\ref{eqJtot}) gives the total current density $J$ (proportional to the electric current in the circuit attached to the superlattice). The restitution coefficients are rescaled as $\alpha_e=\delta\gamma_e$, $\alpha_j=\delta\gamma_j$, where $\delta=1/(\nu[t])\ll 1$ is the ratio between the Bloch period and the convective time scale $[t]$ in table \ref{t1}. The hydrodynamic equations hold in the limit as $\delta\to 0+$  \cite{BAC11}. There are two time scales in the hydrodynamic equations: a nonlinear fast time scale $\theta$ (the phase of the BOs, on the picosecond time scale), given by (\ref{eqtheta}),  and a slow time scale $t$ on the nanosecond scale. The electric field $-F(x,t)$, the electron density $n(x,t)$, the BOs complex amplitude envelope $A(x,t)$ and the $\theta$-averaged current density $\left<J\right>_\theta(t)$ all vary on the slow time scale and may exhibit Gunn-type oscillations with frequencies on the GHz scale. Once we have obtained $F$, $\left<J\right>_\theta$ and $A$ from (\ref{ddeF}) and (\ref{eqA}), which depend only on the slow time scale $t$, we can calculate from (\ref{eqJtot}) the total current density $J$, which depends on both  time scales. Although $A=0$ is an exact solution of (\ref{eqA}), we are interested in finding numerical  solutions with undamped BOs (i.e. $A\neq 0$) coexisting with Gunn type oscillations of the current. The first term in the RHS of (\ref{eqA}) tries to send $A$ to $0$ at a rate proportional to  $(\gamma_{e}+\gamma_{j})/2$, but this effect may be compensated by the second term of the RHS of (\ref{eqA}). This means that nonzero solutions of the  BO amplitude may be present below a critical value of  $(\gamma_{e}+\gamma_{j})$, i.e., when the scattering time is long enough.

\subsection{Boundary conditions}
Equations (\ref{ddeF})-(\ref{poiss}) must be solved together with the following boundary conditions at the cathode ($x=0$):
\begin{eqnarray}
&& \frac{\partial F(0,t)}{\partial t} + \sigma_0 F(0,t) = \left<J\right>_\theta(t),  \nonumber \\
&& A(0,t) = 0, \label{E.bciny1}
\end{eqnarray}
and at the anode ($x=L$):
\begin{eqnarray}
&& \frac{\partial F(L,t)}{\partial t} + \sigma_L n(L,t) F(L,t) = \left<J\right>_\theta(t),  \nonumber \\
&& A(L,t) = 0, \label{E.bciny2}
\end{eqnarray}
together with the voltage bias integral constraint:
\begin{eqnarray}
&& \frac{1}{L}\int_0^L F(x,t)\, dx = \phi, \label{E.bias}
\end{eqnarray}
where $\sigma_0$ and $\sigma_L$ are the dimensionless conductivities at the cathode and anode respectively,  $L$ is the nondimensional SL length and $\phi L$ is the nondimensional constant applied voltage bias.

\begin{figure}
\begin{center}
\includegraphics[width=8cm,angle=0]{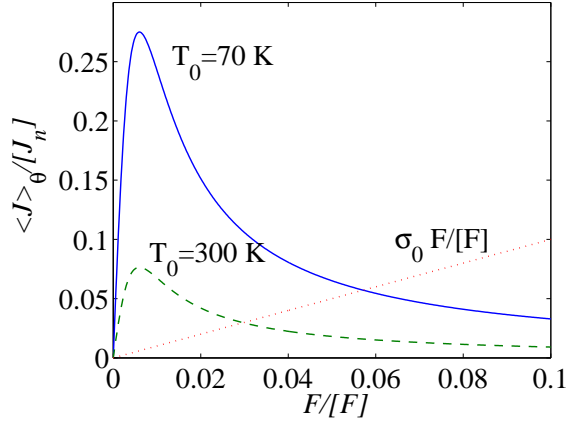}
\vspace{0.2cm} \caption{$\theta$-averaged current density versus electric field for a spatially homogeneous stationary state at different
temperatures.} \label{fig0}
\end{center}
\end{figure}

The contact conductivity at the cathode $\sigma_0$ must be selected so that $\sigma_0 F$ intersects the second branch of the drift velocity $\left<J\right>_\theta(F)$ depicted in Fig. \ref{fig0}. $\left<J\right>_\theta(F)$ is found by solving (\ref{ddeF}) for $F$, provided $n=1$ and $F$ is independent of $x$ and $t$:
\begin{eqnarray}
&& \left<J\right>_\theta(F) = {\delta\gamma_e E_0 F \over F^2+\delta^2\gamma_{j}\gamma_e}. \label{J(F)}
\end{eqnarray}
This is a typical boundary condition that yields Gunn type self-sustained oscillations of the current in drift-diffusion SL models
\cite{BT10,BEP03,b02,pbe04}.

\subsection{Initial conditions}
We select spatially uniform $A$ and $F$ as initial conditions:
\begin{eqnarray}
&& A(x,0) = \frac{I_1\!\left(\frac{\Delta}{2 k_B T_0}\right)}{ I_0\!\left(\frac{\Delta}{2 k_B T_0}\right)} ,\\
&& F(x,0) = \phi,
\end{eqnarray}
where $I_j(x)$ is the modified Bessel function \cite{BAC11}. The initial value of $A$ is calculated by using (\ref{EqNewton1}) assuming that the initial values of $u^{(0)}$, $\beta^{(0)}$ and $\theta$ are 0, $\Delta/(2 k_B T_0)$ and $0$, respectively. The initial nondimensional mean energy is the lattice mean energy density:
\begin{eqnarray}
&& E_0=\frac{I_1\!\left(\frac{\Delta}{2k_B T_0}\right)}{I_0\!\left(\frac{\Delta}{2k_BT_0} \right)}. \label{E0}
\end{eqnarray}

 Although the initial-boundary value problem (IBVP) to be solved is one dimensional, we need efficient numerical methods because it is very computationally intensive: we need a large integration time to observe the Gunn-type self-oscillations of the current and the  number of operations needed to compute the double Fourier coefficients $f^{B(0)}_{2,0}$, $f^{B(0)}_{2,-1}$, $f^{B(0)}_{3,-1}$ and $f^{B(1)}_{2,-1}$ given by (\ref{fBj})-(\ref{eq93}) is large. The numerical computation of the multipliers $u$ and $\beta$ (which depend on both fast and slow time scales) is one of the bottlenecks of this problem. To calculate $f^{B(0)}_{2,0}$, $f^{B(0)}_{2,-1}$ and $f^{B(0)}_{3,-1}$ in (\ref{ddeF}) and (\ref{eqA}), we need to know $u^{(0)}$ and $\beta^{(0)}$ given by solving (\ref{eqNewton}). To calculate $f^{B(1)}_{2,-1}$ we need $u^{(1)}$ and $\beta^{(1)}$ given by (\ref{equ1beta1}).

\section{Numerical solution} \label{sec:3}
After replacing $n=1+\partial F/\partial x$ from (\ref{poiss}), the drift-diffusion equation (\ref{ddeF}) can be written in the following
way:
\begin{eqnarray}
&& \frac{\partial F}{\partial t} + \mathcal A \,\frac{\partial F}{\partial x} + \mathcal B\, \frac{\partial^2 F}{\partial x^2} +
\mathcal{C}\, \left<J\right>_\theta =  \mathcal D,\label{E.dF/dt}
\end{eqnarray}
where coefficients $\mathcal A$, $\mathcal B$, $\mathcal C$ and $\mathcal D$ are:
\begin{eqnarray}
&& \mathcal A = {\delta\gamma_e E_0 F \over F^2+\delta^2\gamma_{j}\gamma_e} \label{coefA} \\
&& \mathcal B = -{\delta^2\gamma_e  \over 2(F^2+\delta^2\gamma_{j}\gamma_e)} \label{coefB}\\
&& \mathcal C = -1 + {\delta^3\gamma_e E_0  \over (F^2+\delta^2\gamma_{j}\gamma_e)^3}\left[2 \gamma_j F^2
+ \gamma_e(F^2-\delta^2\gamma_{j}\gamma_e)\right]\left( 1+{\partial F \over \partial x}\right) \label{coefC} \\
&& \mathcal D =  J_{n,Su}\,{\mathcal C}  -{\delta \over F^2+\delta^2\gamma_{j}\gamma_e}\left[ {F \over 2} {\partial \over \partial x}
\mbox{Im}\,\left({f_{2,0}^{B(0)} \over 1+2iF} \right)  + {\delta \gamma_e \over 2} {\partial \over \partial
x} \mbox{Re}\,\left({f_{2,0}^{B(0)} \over 1+2iF} \right)\right. \nonumber\\
&&\quad\quad + \left. \left({\delta^2 \gamma_e (\gamma_e + \gamma_j) E_0 F \over F^2+\delta^2\gamma_{j}\gamma_e}\right) {\partial J_{n,Su}
\over \partial x} \right]   \label{coefD}
\end{eqnarray}
The influence of the amplitude $A$ in equation (\ref{ddeF}) is small because only the term $f^{B(0)}_{2,0}$ in (\ref{coefD}) contains it. On the other hand, in equation (\ref{eqA}) we can separate the diffusion term from the rest of the terms. Note that the amplitude $A$ appears implicitly in the double Fourier coefficients $f_{2,-1}^{B(0)}$, $f_{3,-1}^{B(0)}$ and $f_{2,-1}^{B(1)}$. Therefore, we can write (\ref{eqA}) as follows:
\begin{eqnarray}
&&\frac{\partial A}{\partial t}=-\frac{\gamma_{e}+\gamma_{j}}{2} \,A + \frac{1}{2i}\frac{\partial}{\partial
x}\!\left(\frac{f^{B(0)}_{2,-1} + \delta\, s^{(1)}_{2,-1}}{1+iF}\right) + \frac{\delta}{4(1+iF)}\frac{\partial^2 A}{\partial x^2} \nonumber\\
&&\quad\quad -\frac{\delta i}{4(1+iF)^2}\frac{\partial A}{\partial x}\frac{\partial F}{\partial x}  \!,  \label{eqA1}\\
&&s^{(1)}_{2,-1} = f^{B(1)}_{2,-1}\! -\!\left(\mathcal{A}^{(0)} \frac{\partial}{\partial A}+(\langle
J\rangle_\theta-J_{n,Su})\frac{\partial}{\partial F}-\frac{\partial J_{n,Su}}{\partial x}\frac{\partial}{\partial
n}\right)\!\frac{f^{B(0)}_{2,-1}}{1+iF}\nonumber\\
&&\quad\quad +\frac{1}{2i}\frac{\partial}{\partial x}\!\left(\frac{f^{B(0)}_{3,-1}}{1+2iF}\right)\!. \nonumber
\end{eqnarray}
For the numerical computation of $u^{(0)}$ and $\beta^{(0)}$ we can write equation (\ref{eqNewton})  as:
 \begin{eqnarray}
&&f^{B(0)}_{1} = n\,G(u^{(0)},\beta^{(0)})=  A\, e^{-i\theta}, \label{EqNewton1}
\end{eqnarray}
in which
 \begin{eqnarray}
&&  G(u^{(0)},\beta^{(0)})= \frac{K_{c\beta}  - i K_s}{K_c}. \label{E.Gubeta}
\end{eqnarray}
where
 \begin{eqnarray}
&&K_c =   \int_{0}^\pi e^{\beta^{(0)}\cos k}\cosh(u^{(0)}k)\, dk \label{Int1}\\
&&K_s = \int_{0}^\pi e^{\beta^{(0)}\cos k}\sinh(u^{(0)}k)\sin k\, dk  \label{Int2}\\
&&K_{c\beta} = {\partial K_c \over \partial \beta^{(0)} } = \int_{0}^\pi e^{\beta^{(0)}\cos k}\cosh(u^{(0)}k)\cos k\, dk  \label{Int3}
\end{eqnarray}
For an efficient numerical computation of integrals (\ref{Int1})-(\ref{Int3}) we use an  expansion as  series of modified Bessel functions
described in \ref{app}. Once we have obtained $u^{(0)}$ and $\beta^{(0)}$ from (\ref{EqNewton1}), we can get $u^{(1)}$ and $\beta^{(1)}$ as shown in Appendix \ref{app}.

\subsection{Numerical scheme} We have used an implicit numerical scheme to solve the  partial differential equations (\ref{ddeF}) and (\ref{eqA}). In order to avoid numerical instabilities, our scheme employs a fixed point iteration method to find the numerical solution for $F$, $A$, $\left<J_\theta \right>$ and $J$ at each time step.

\subsubsection{Drift-diffusion equation for $F$} To solve equation (\ref{ddeF}), we use a scheme similar to the one described and proved to converge in \cite{CHK1} for a related problem involving partial differential equations with an integral constraint. We use central differences for approximating spatial derivatives, and the
resulting differential equation is integrated in time by a first order implicit Euler method. This procedure leads to a system of $N+2$ linear equations for the $N+1$ values of the electric field $F_j^{\rm n}\approx F(jh,{\rm n}\tau)$, $j=0,1,\dots,N$, with the subscript $j$
referring to space and the superscript ${\rm n}$ to time, plus  $\left<J\right>_\theta$. In order to save computational effort we will set up the finite difference system of equations with a tridiagonal coefficients matrix, in the following way:
\begin{eqnarray}
& \mathit{a}_{j}F_{j-1}^{{\rm n}+1} + \mathit{b}_{j}F_{j}^{{\rm n}+1} + \mathit{c}_{j}F_{j+1}^{{\rm n}+1} +
\mathit{d}_{j}\left<J\right>_\theta^{{\rm n}+1} = \mathit{g}_{j}, \,\,\, &j=1,\dots,N-1 \label{E.finitedif}
\end{eqnarray}
The coefficients of (\ref{E.finitedif})  are:
\begin{eqnarray}
&& \mathit{a}_{j} = -{h\tau \over 2}\,\mathcal A_{j}^{{\rm n}+1} + \tau\,\mathcal B_{j}^{{\rm n}+1} \label{coefaj} \\
&& \mathit{b}_{j} = h^2 - 2\tau\, \mathcal B_{j}^{{\rm n}+1}  \nonumber \\
&& \mathit{c}_{j} = {h\tau \over 2}\,\mathcal A_{j}^{{\rm n}+1} + \tau\,\mathcal B_{j}^{{\rm n}+1} \nonumber \\
&& \mathit{d}_{j} = h^2\tau\,\mathcal C_{j}^{{\rm n}+1}  \nonumber \\
&& \mathit{g}_{j} = h^2\, F_j^{\rm n} +  h^2\tau\, \mathcal D_{j}^{{\rm n}+1} ,  \label{coefgj}
\end{eqnarray}
where $h=\Delta x=L/N$, $\tau=\Delta t$ and  all the coefficients $\mathcal A$, $\mathcal B$, $\mathcal C$ and $\mathcal D$ of
(\ref{E.dF/dt}) are evaluated at time $t^{{\rm n}+1}$.\\
The voltage bias integral constraint is solved by the composite Simpson's rule:
\begin{eqnarray}
&& F_0^{{\rm n}+1} + 4F_1^{{\rm n}+1} + 2F_2^{{\rm n}+1} + ... + 2F_{N-2}^{{\rm n}+1} + 4F_{N-1}^{{\rm n}+1} + F_{N}^{{\rm n}+1} = 3\phi L
/h. \label{E.Simpson1}
\end{eqnarray}
The boundary condition at the injector contact is:
\begin{eqnarray}
&& (1+\sigma_0\,\tau)F_{0}^{{\rm n}+1} - \tau\, \left<J\right>_\theta^{{\rm n}+1} = F_0^{\rm n}, \label{C.bcinynum}
\end{eqnarray}
and at the collector contact:
\begin{eqnarray}
&& ( h + \sigma_L \tau(h + F_{N}^{{\rm n}+1} - F_{N-1}^{{\rm n}+1}))F_{N}^{{\rm n}+1} -\tau h\left<J\right>_\theta^{{\rm n}+1} =
hF_{N}^{{\rm n}}. \label{C.bccolnum}
\end{eqnarray}
This system of $N+2$ linear equations can be reduced to a simpler and smaller system, with the objective of finding a tridiagonal
matrix, in the following way:
\begin{itemize}
  \item $\left<J\right>_\theta$ can be calculated directly from the boundary condition at the injector contact:
  \begin{eqnarray}
     && \left<J\right>_\theta^{{\rm n}+1} = (\sigma_0 + {1 \over \tau})F_{0}^{{\rm n}+1} - {1 \over  \tau} F_0^{\rm n}.  \label{E.Jn+1}
  \end{eqnarray}
  \item The field at the anode can also be expressed in terms of $F_{0}^{{\rm n}+1}$ and $F_{N-1}^{{\rm n}+1}$:
  \begin{eqnarray}
&&  F_{N}^{{\rm n}+1} = {h(F_{N}^{{\rm n}}-F_{0}^{{\rm n}} + (1+\sigma_0 \tau)F_{0}^{{\rm n}+1})  \over h + \sigma_L \tau (h + F_{N}^{{\rm
n}+1} - F_{N-1}^{{\rm n}+1})}. \label{E.FN}
\end{eqnarray}
\item We can make the following factorization of the system of linear equations:
  \begin{eqnarray}
    && \textbf{v}\, F_0^{{\rm n}+1} + T\, \textbf{F} = \textbf{g}      \label{E.vF0TFg} \\
   && (1+\kappa_1)F_0^{{\rm n}+1} + \textbf{u}\cdot\textbf{F} = 3\phi L/h - \kappa_2,  \label{E.F0uF}
  \end{eqnarray}
  where coefficients $\kappa_1$ and $\kappa_2$ are:
\begin{eqnarray}
    && \kappa_1 = {h(1+\sigma_0 \tau) \over h + \sigma_L \tau (h+ F_{N}^{{\rm n}+1} - F_{N-1}^{{\rm n}+1})}, \nonumber \\
    && \kappa_2 = {h(F_{N}^{{\rm n}}-F_0^{\rm n}) \over h + \sigma_L \tau  (h+ F_{N}^{{\rm n}} - F_{N-1}^{{\rm n}+1} )}, \nonumber
  \end{eqnarray}
 $T$ is the tridiagonal matrix:
 $$ T = \left(
    \begin{array}{ccccc}
      \mathit b_1 & \mathit c_1 & \cdots & \cdots & 0 \\
      \mathit a_2 & \mathit b_2 & \mathit c_2 & \cdots & 0 \\
      \cdots & \cdots & \cdots & \cdots & \cdots \\
      0 & \cdots & \cdots & \mathit a_{N-1} & \mathit b_{N-1} \\
    \end{array}
  \right) $$
  and  vectors $\textbf{F}$, $\textbf{v}$, $\textbf{g}$ and $\textbf{u}$ are:
  $$\textbf{F} = \left(
                   \begin{array}{c}
                     F_1^{{\rm n}+1} \\
                     F_2^{{\rm n}+1} \\
                     \cdots \\
                     F_{N-1}^{{\rm n}+1} \\
                   \end{array}
                 \right), \quad
    \textbf{v} = \left(
                   \begin{array}{l}
                     \mathit d_1(\sigma_0+{1 \over \tau})+\mathit a_1  \\
                     \mathit d_2(\sigma_0+{1 \over \tau})  \\
                     \cdots \cdots \cdots \cdots \\
                     \mathit d_{N-1}(\sigma_0+{1 \over \tau}) + \mathit c_{N-1}\, \kappa_1  \\
                   \end{array}
                 \right), $$  \\
   $$  \textbf{g} = \left(
                   \begin{array}{c}
                    \mathit g_1 + {1 \over \tau} F_0^{\rm n} \mathit d_1  \\
                    \mathit g_2 + {1 \over \tau} F_0^{\rm n} \mathit d_2  \\
                     \cdots \cdots \\
                    \mathit g_{N-1} + {1 \over \tau} F_0^{\rm n} \mathit d_{N-1}  \\
                   \end{array}
                 \right), \quad
   \textbf{u} = \left(4, \quad  2,\,  \ldots \, 2, \quad  4 \right).
   $$

System (\ref{E.vF0TFg})-(\ref{E.F0uF}) can be efficiently solved by means of the following system with the same tridiagonal
matrix $T$:
\begin{eqnarray}
    && T\,\textbf{y} = \textbf{g}      \label{E.Tyg} \\
    && T\,\textbf{z} = \textbf{v}      \label{E.Tzv}
  \end{eqnarray}
After calculating $\textbf{y}$ and $\textbf{z}$, we can obtain $F_0^{{\rm n}+1}$, $\textbf{F}$ and $\left<J\right>_\theta^{{\rm n}+1}$:
\begin{eqnarray}
    && F_0^{{\rm n}+1} = {\textbf{u}\cdot\textbf{y}-3\phi L/h + \kappa_2 \over \textbf{u}\cdot\textbf{z}-1-\kappa_1}, \label{E.F0res} \\
    && F_N^{{\rm n}+1} = \kappa_1 F_0^{{\rm n}+1} + \kappa_2,  \\
    && \textbf{F}      = \textbf{y} - F_0^{{\rm n}+1} \textbf{z},                                 \label{E.Fres} \\
    && \left<J\right>_\theta^{{\rm n}+1} = (\sigma_0 + {1 \over \tau}) F_0^{{\rm n}+1} - {1 \over \tau} F_0^{{\rm n}}. \label{E.Jn+1res}
  \end{eqnarray}
\item For each $\theta\in[-\pi, \pi]$, the nondimensional multipliers $\beta_j^{{\rm n}+1,\theta}$ and $u_j^{{\rm n}+1,\theta}$ are obtained by solving (\ref{E.Gubeta}) using  the Newton-Raphson method with the Jacobian matrix (\ref{Jacobian}). The Boltzmann distribution function Fourier coefficients $\left(f^{B(0)}_{2,0}\right)_j^{{\rm n}+1}$, $\left(f^{B(0)}_{2,-1}\right)_j^{{\rm n}+1}$ and $\left(f^{B(0)}_{3,-1}\right)_j^{{\rm n}+1}$ are calculated from (\ref{fBj})-(\ref{fBjl}) using the composite Simpson's rule for all the integrals over $k$ and over $\theta$.
\end{itemize}

\subsubsection{Parabolic equation for $A$.} It is important to find an accurate finite difference scheme to solve (\ref{eqA}) because the restitution parameters $\gamma_e,\gamma_j$ are close to their critical values; thus we will use a second-order implicit scheme. We approximate the time derivative of the complex amplitude $A$ at time ${\rm n}+1$ as:
\begin{eqnarray}
\frac{\partial A}{\partial t}(x_j,t^{{\rm n}+1}) = \frac{\partial A_j^{{\rm n}+1}}{\partial t} = \frac{2}{\tau}\left(A_j^{{\rm
n}+1}-A_j^{{\rm n}} \right) - \frac{\partial A_j^{{\rm n}}}{\partial t} + O(\tau^2);
\end{eqnarray}
which is equivalent to a second-order, implicit, Runge-Kutta method (trapezoidal rule). We use central difference for  the space derivative
in the second term of the RHS of (\ref{eqA1}) as a whole block:
\begin{eqnarray}
\frac{\partial}{\partial x}\!\!\left(\!\frac{f^{B(0)}_{2,-1} + \delta\, s^{(1)}_{2,-1}}{1+iF}\!\right)\!\approx
\frac{1}{2h}\!\left[\!\frac{\!\left(\!f^{B(0)}_{2,-1} + \delta s^{(1)}_{2,-1}\! \right)_{j+1}^{{\rm n}+1}}{1+iF_{j+1}^{{\rm n}+1}} -
\frac{\!\left(f^{B(0)}_{2,-1} + \delta\, s^{(1)}_{2,-1}\right)_{j-1}^{{\rm n}+1}}{1+iF_{j-1}^{{\rm n}+1}}\!\right]\!\label{2diff}
\end{eqnarray}
Therefore, the numerical scheme for $A$ will be as follows:
\begin{eqnarray}
&& A_j^{{\rm n}+1} = A_j^{{\rm n}} + {\tau \over 2}\frac{\partial A_j^{{\rm n}}}{\partial t} - {\tau \over 4}(\gamma_{e} + \gamma_{j})\,A_j^{{\rm n}+1}   \nonumber \\
&& \quad + \frac{\tau}{8ih}\left( \frac{\left(f^{B(0)}_{2,-1} + \delta\, s^{(1)}_{2,-1} \right)_{j+1}^{{\rm n}+1}}{1+iF_{j+1}^{{\rm
n}+1}} - \frac{\left(f^{B(0)}_{2,-1} + \delta\, rs^{(1)}_{2,-1}\right)_{j-1}^{{\rm n}+1}}{1+iF_{j-1}^{{\rm n}+1}} \right) \nonumber\\
&& \quad + \frac{\tau \delta\, (A_{j+1}^{{\rm n}+1} -2A_{j}^{{\rm n}+1} + A_{j-1}^{{\rm n}+1})}{8(1+iF)h^2}
 -  \frac{\tau \delta i\, (A_{j+1}^{{\rm n}+1} - A_{j-1}^{{\rm n}+1})(F_{j+1}^{{\rm n}+1} - F_{j-1}^{{\rm n}+1})}{32(1+iF)^2 h^2}   \label{E.Aexplicit}
\end{eqnarray}
where $j=1,\dots,N-1$. The boundary conditions at the  contacts are:
\begin{eqnarray}
&& A_0^{{\rm n}+1} = 0, \\
&& A_N^{{\rm n}+1} = 0. \label{E.AN}
\end{eqnarray}

\subsubsection{Algorithm}
\begin{enumerate}
\item For each time step $t^{{\rm n}+1}$ do:
\begin{enumerate}
\item While the fixed point iteration does not converge:
\begin{itemize}
\item For each node $j=1\dots N-1$:
\begin{itemize}
\item For each $\theta\in[-\pi, \pi]$: calculate  the Jacobian matrix elements  from (\ref{Jacobian}) using the Bessel functions series
described in Appendix \ref{app} for the integrals (\ref{Int1})-(\ref{Int3}) and (\ref{Int4})-(\ref{Int8}), and then obtain $(\beta^{(0)})_j^{{\rm
n}+1,\theta}$ and $(u^{(0)})_j^{{\rm n}+1,\theta}$ by the Newton-Raphson method from (\ref{EqNewton1}).
 \item Calculate $\left(f^{B(0)}_{2,0}\right)_j^{{\rm n}+1}$, $\left(f^{B(0)}_{2,-1}\right)_j^{{\rm n}+1}$ and
$\left(r^{(1)}_{2,-1}\right)_j^{{\rm n}+1}$ from (\ref{fBj}), (\ref{fBjl}) and (\ref{r2m1}). The integrals over $k$ and $\theta$ are
calculated by the composite Simpson's rule, with the previous values obtained of $(\beta^{(0)})_j^{{\rm n}+1,\theta}$ and
$(u^{(0)})_j^{{\rm n}+1,\theta}$. 
 \item Calculate coefficients $\mathcal A_j^{{\rm n}+1}$, $\mathcal B_j^{{\rm n}+1}$, $\mathcal C_j^{{\rm n}+1}$  and $\mathcal D_j^{{\rm n}+1}$ from (\ref{coefA})-(\ref{coefD}).
 \item Calculate coefficients $\mathit{a}_j$, $\mathit{b}_j$, $\mathit{c}_j$ and $\mathit{g}_j$ from (\ref{coefaj})-(\ref{coefgj}).
\end{itemize}
\end{itemize}
\item Obtain $F_j^{{\rm n}+1}$ and $\left<J\right>_\theta^{{\rm n}+1}$ from (\ref{E.F0res})-(\ref{E.Jn+1res}),
      the complex amplitude $A_j^{{\rm n}+1}$ from (\ref{E.Aexplicit})-(\ref{E.AN}) and the current density $J$ from (\ref{eqJtot}).
\item If the fixed point iteration does not converge, then go to  step (a).
\end{enumerate}
\item Go to  step (1).
\end{enumerate}

\subsection{Numerical results}
We have solved the hydrodynamic equations using parameter values similar to those for the superlattice with $\Delta= 16$ meV in Ref.\ \cite{sch98}. Then $\delta=0.0053$ and other parameter values are as in Table \ref{t1}. To obtain undamped BOs, we have used $\gamma_{e,j}=1.1269$ so that $(\gamma_e+\gamma_j)/2<\gamma_{\rm crit}$. We consider a 50-period dc voltage biased GaAs-AlAs SL with lattice temperature 70 K and dimensionless contact conductivities $\sigma_{0}=1$ and $\sigma_L=0.25$. Initially, the mean energy density is $E_0=0.5501$ and the profiles of $A$ and $F$ are uniform, taking on values of $0.5501$ and $0.05$, respectively.\\

\begin{figure}
\begin{center}
\includegraphics[width=8cm,angle=0]{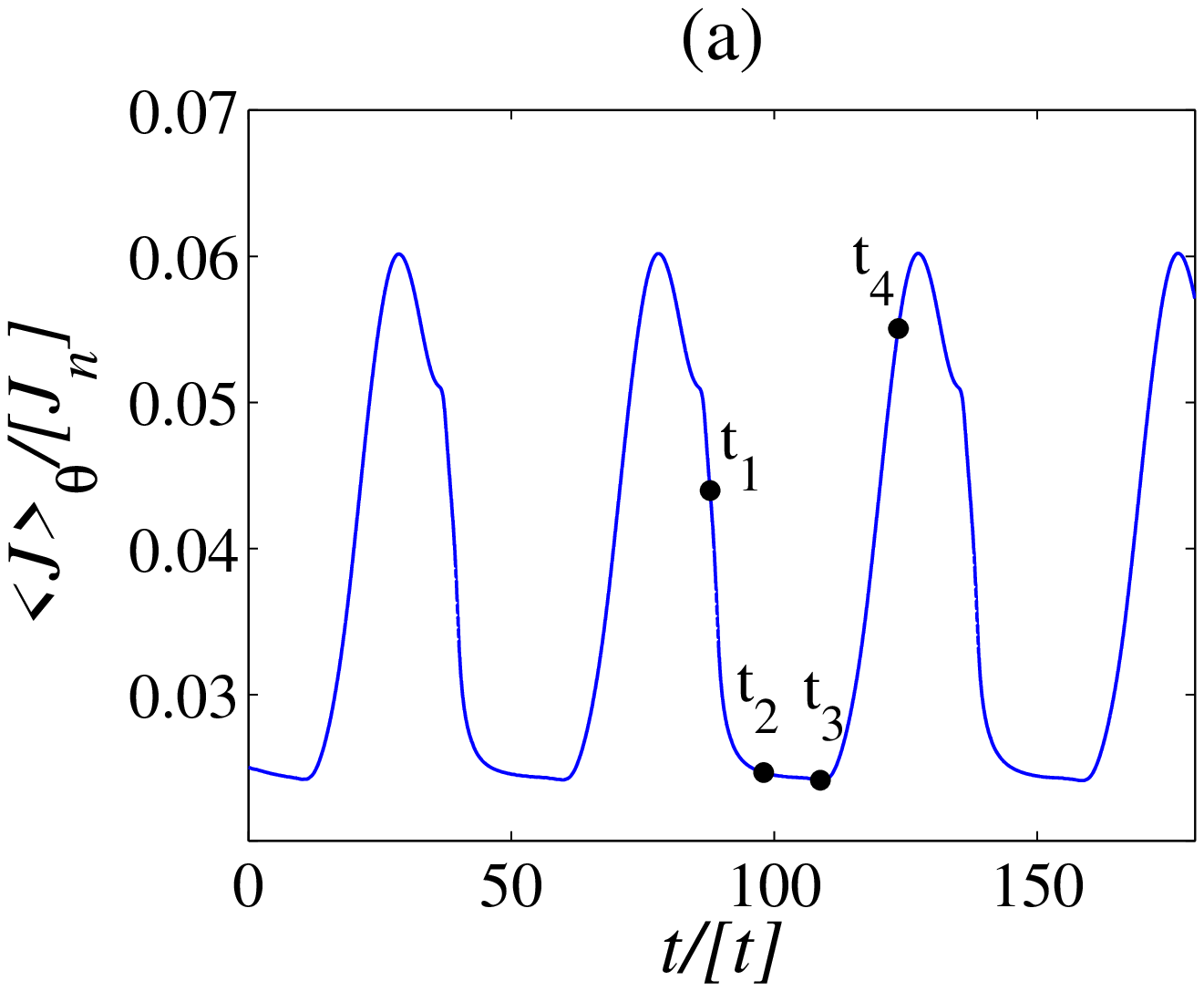}\\
\includegraphics[width=10cm,angle=0]{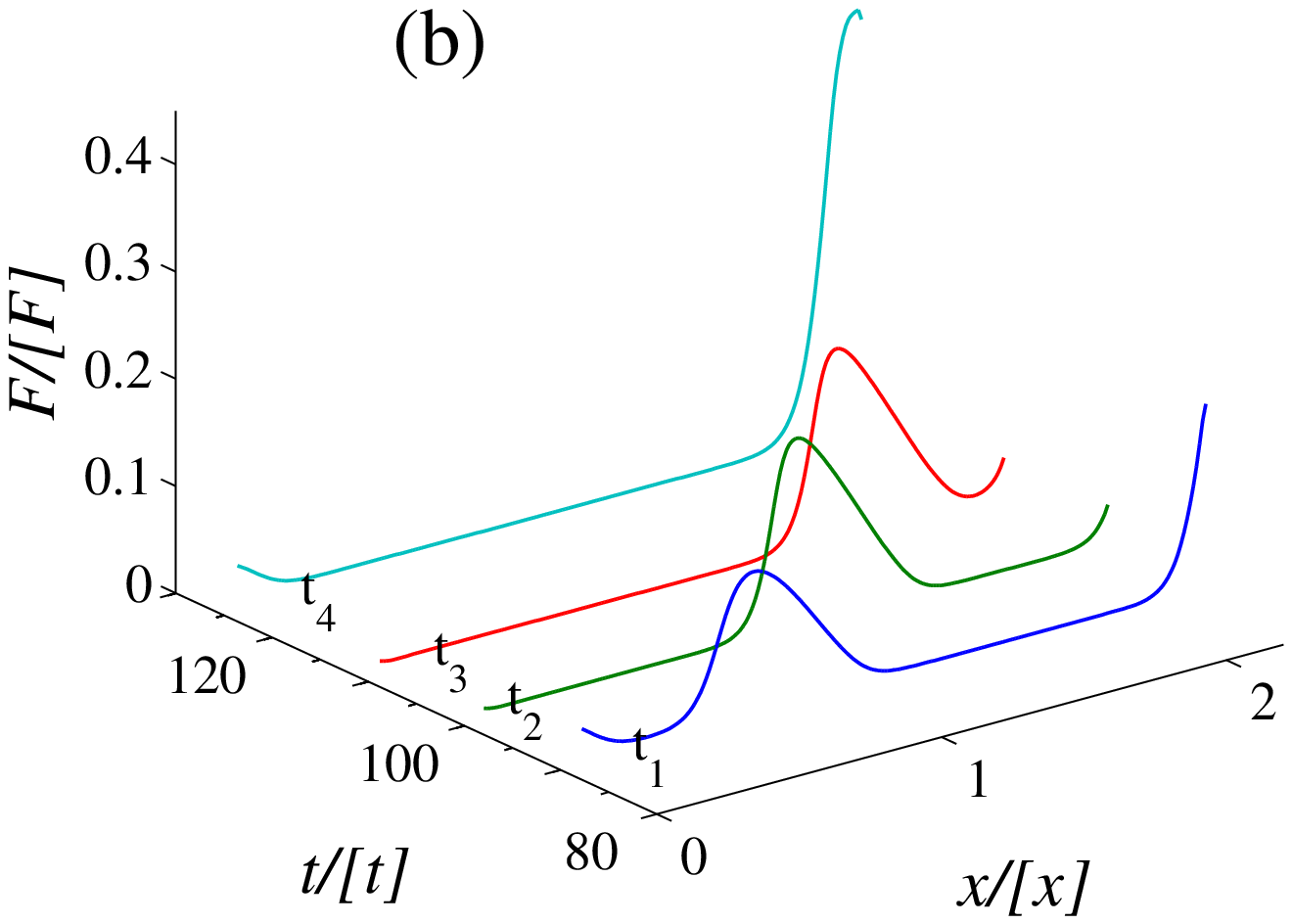} 
\includegraphics[width=10cm,angle=0]{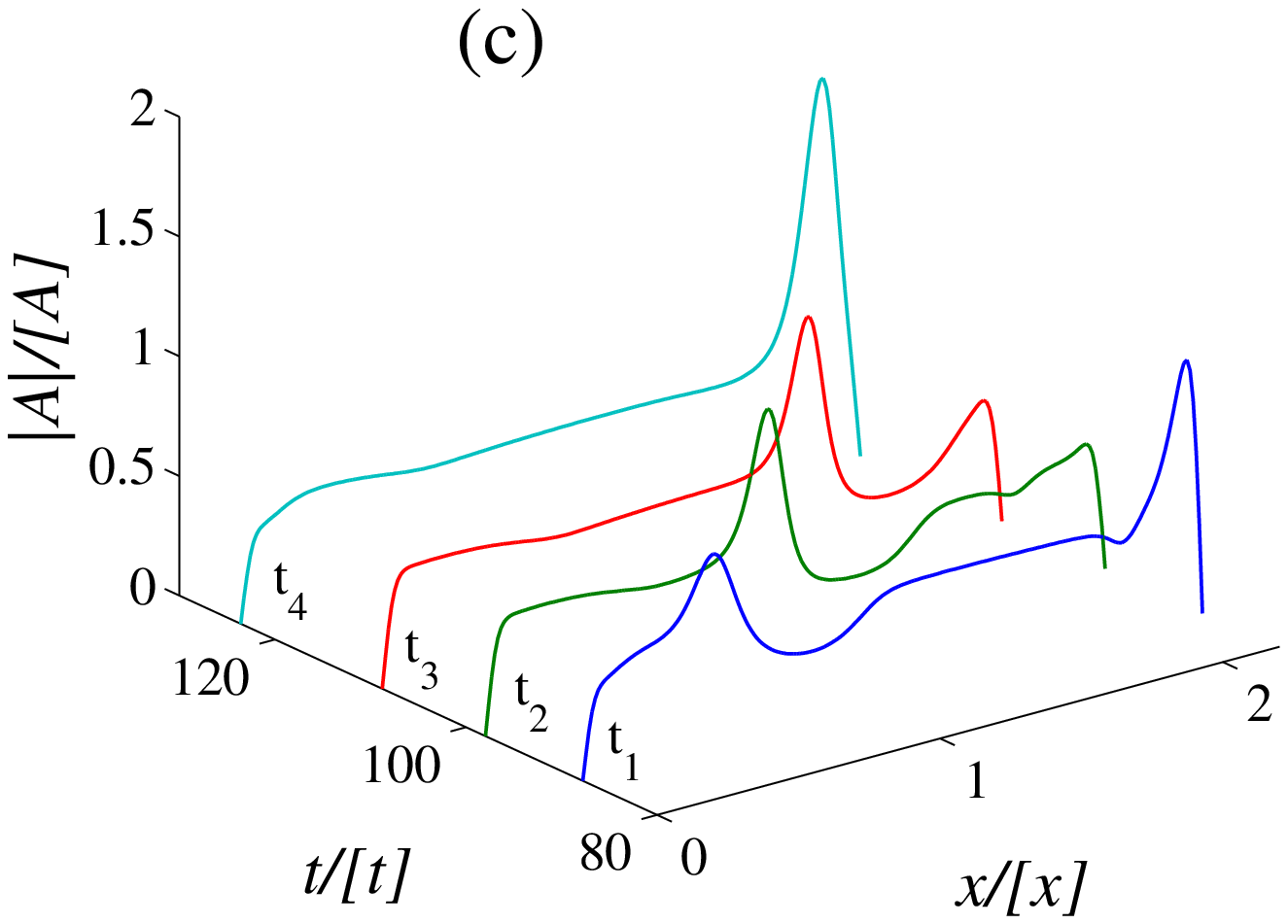}
\end{center}
\begin{center}
\vspace{0.2cm} \caption{(a) $\theta$-averaged total current density vs time during coexisting Bloch and Gunn type oscillations at 70 K. (b)
Field profile vs space at the times $t_1$ to $t_4$ marked in (a).  (c) Same for the complex BO amplitude profile. To transform the
magnitudes in this figure to dimensional units, use Table \ref{t1}.} \label{fig1}
\end{center}
\end{figure}

\begin{figure}
\begin{center}
\includegraphics[width=8cm,angle=0]{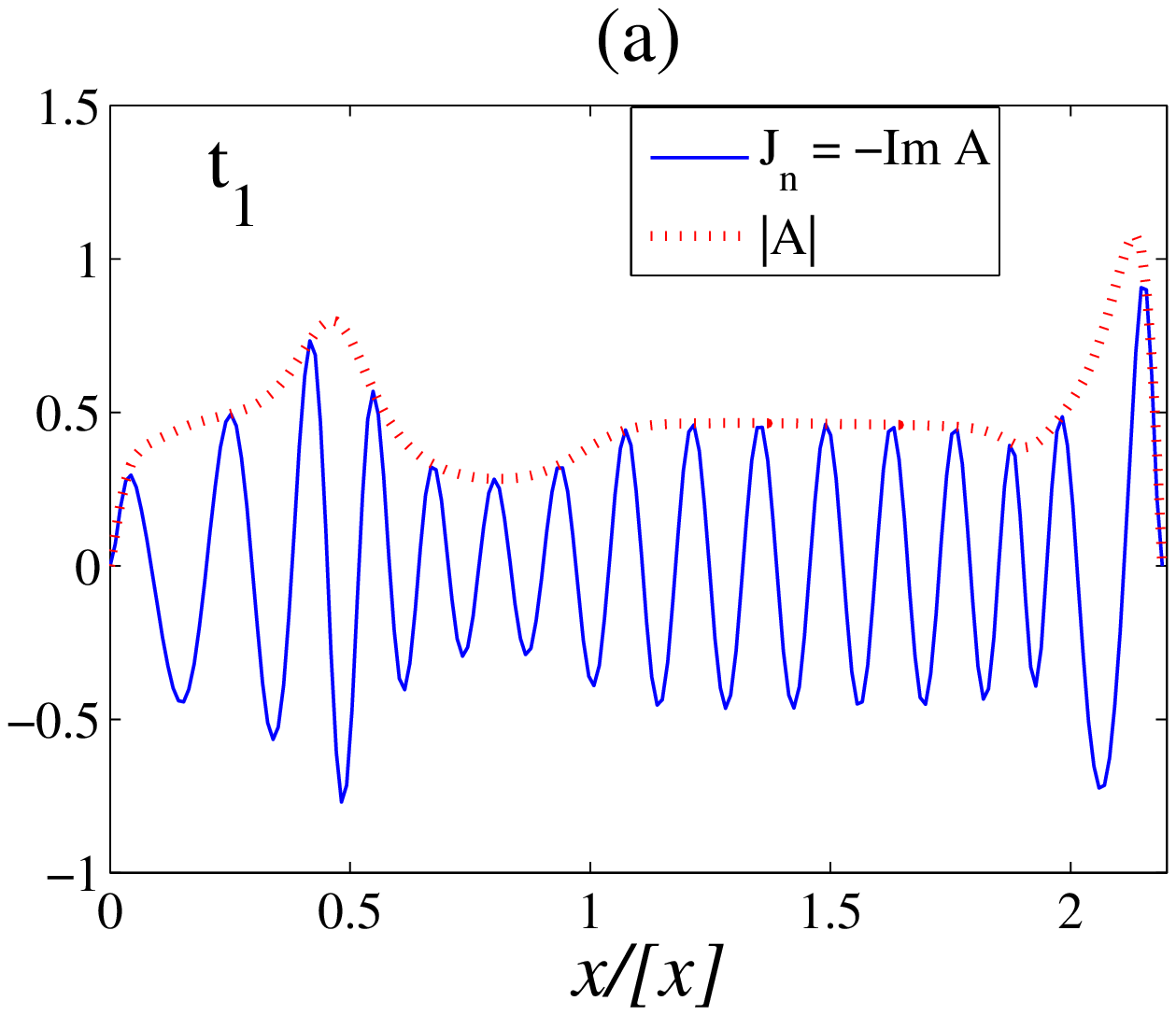}
\includegraphics[width=8cm,angle=0]{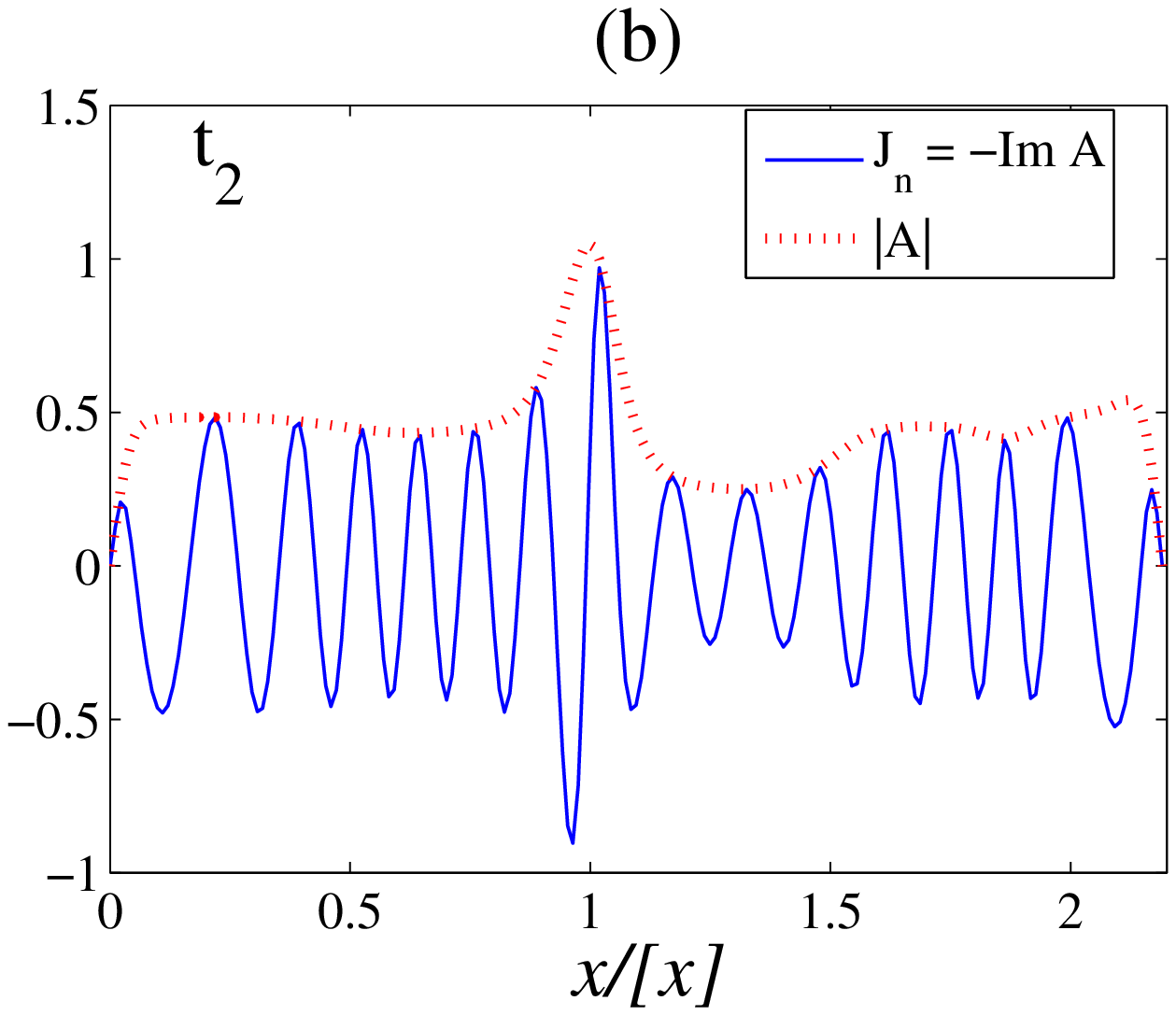}
\end{center}
\begin{center}
\includegraphics[width=8cm,angle=0]{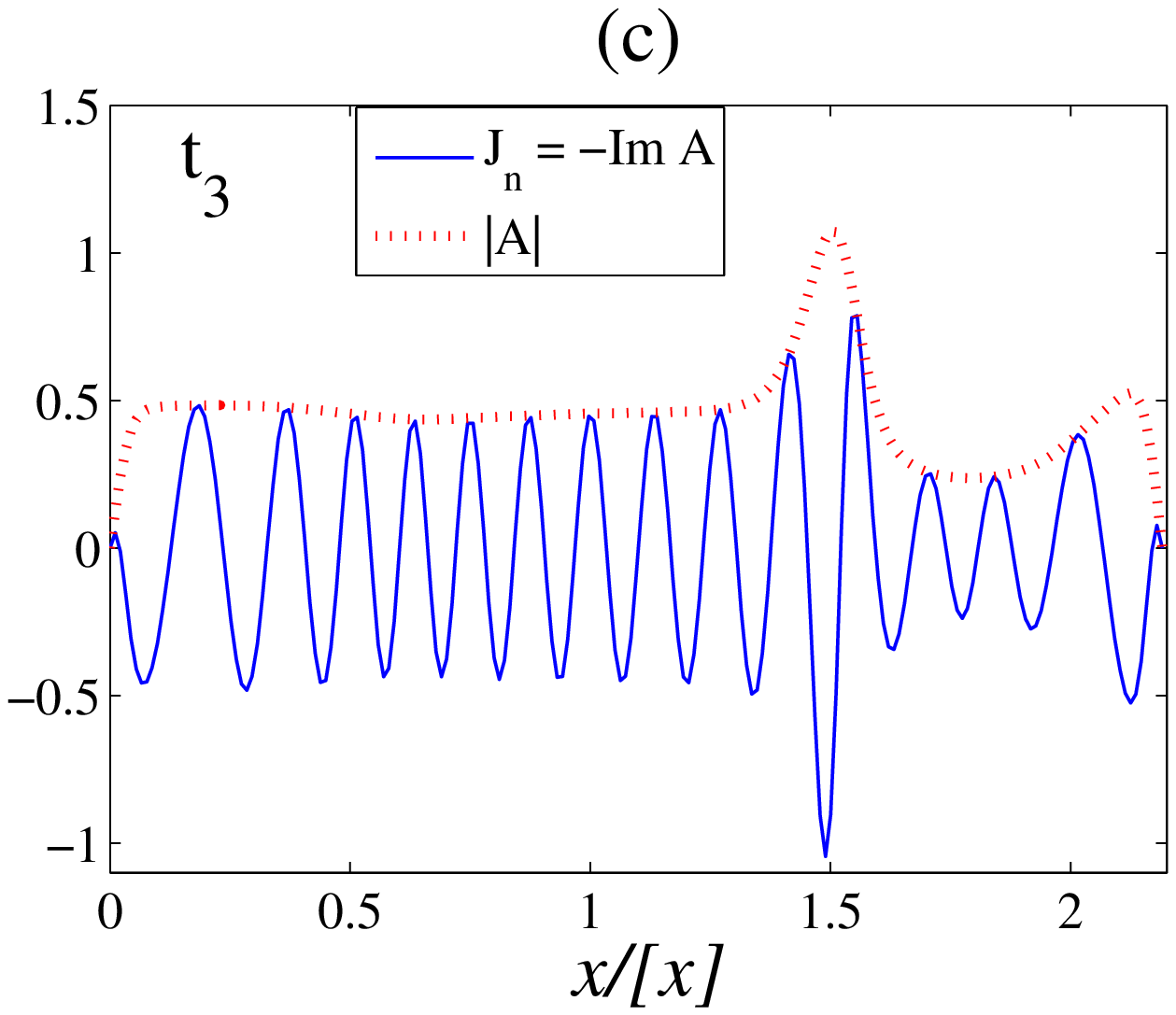}
\includegraphics[width=8cm,angle=0]{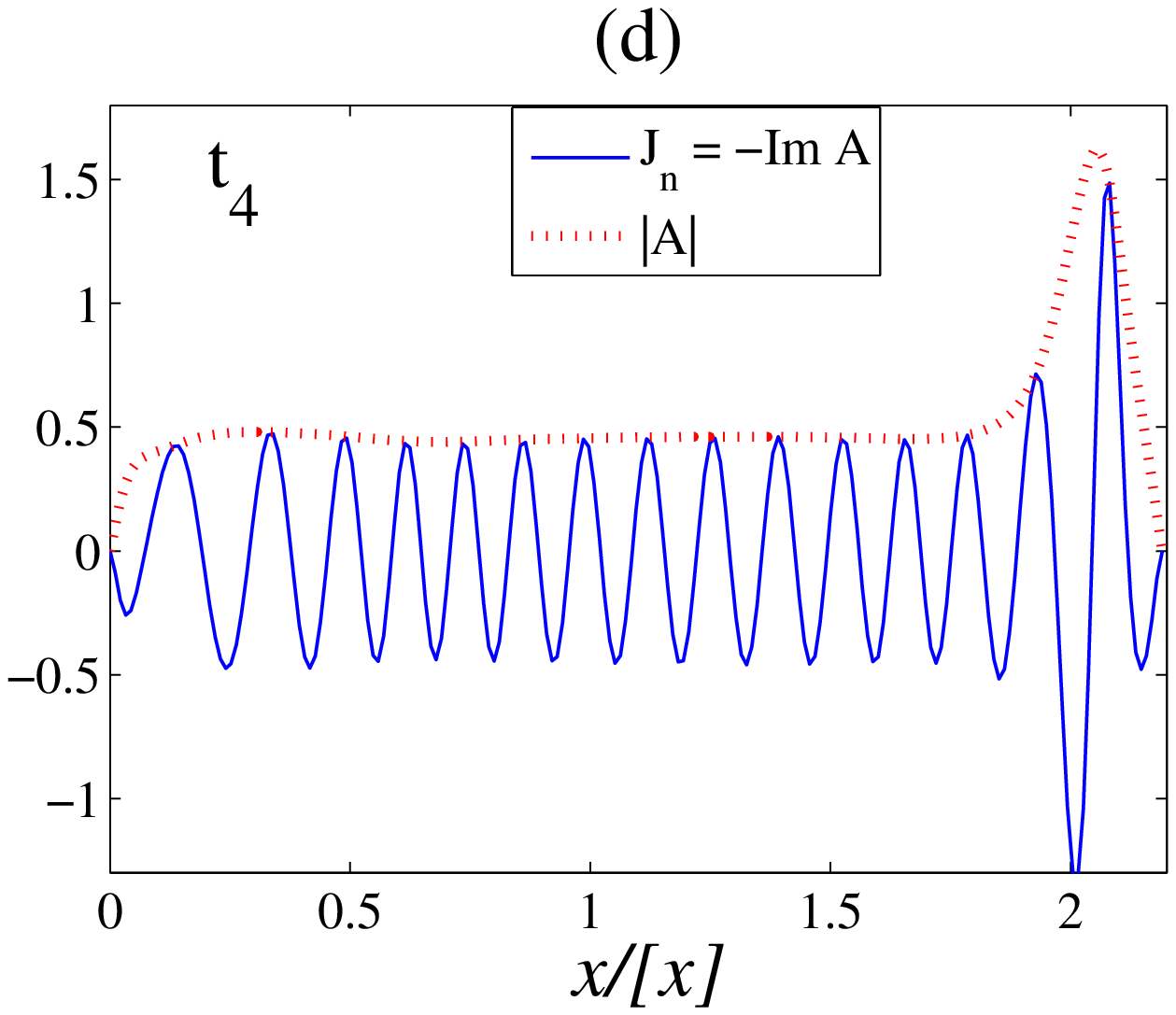}
\end{center}
\begin{center}
\vspace{0.2cm} \caption{Profiles of the nondimensional electron current density $J_n$ and the modulus of the amplitude  for BOs at $\theta
= 0$ corresponding to the instants $t_1-t_4$ of Fig. \ref{fig1}(a).} \label{fig1.1}
\end{center}
\end{figure}

\begin{figure}
\begin{center}
\includegraphics[width=8cm,angle=0]{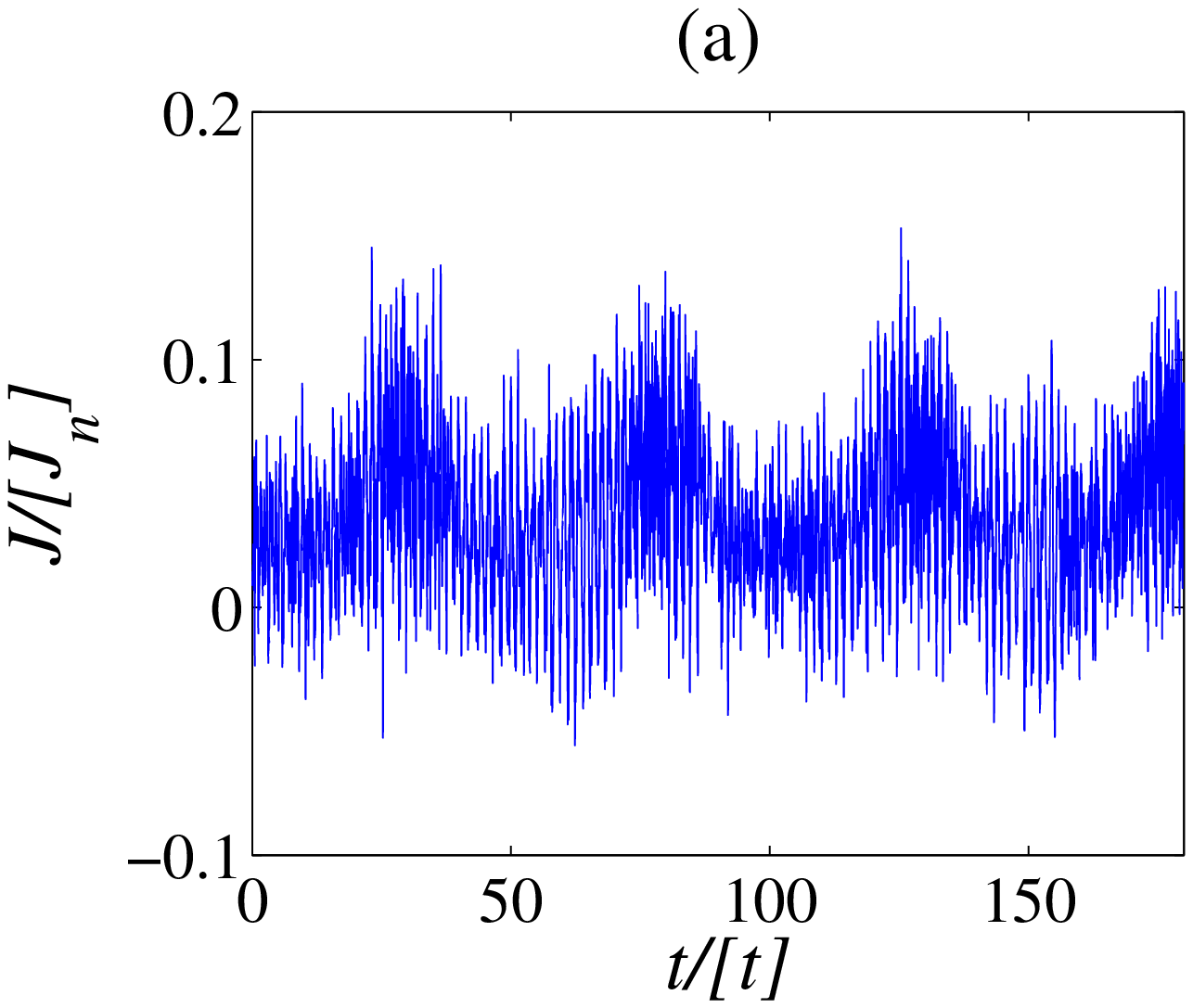}
\includegraphics[width=8cm,angle=0]{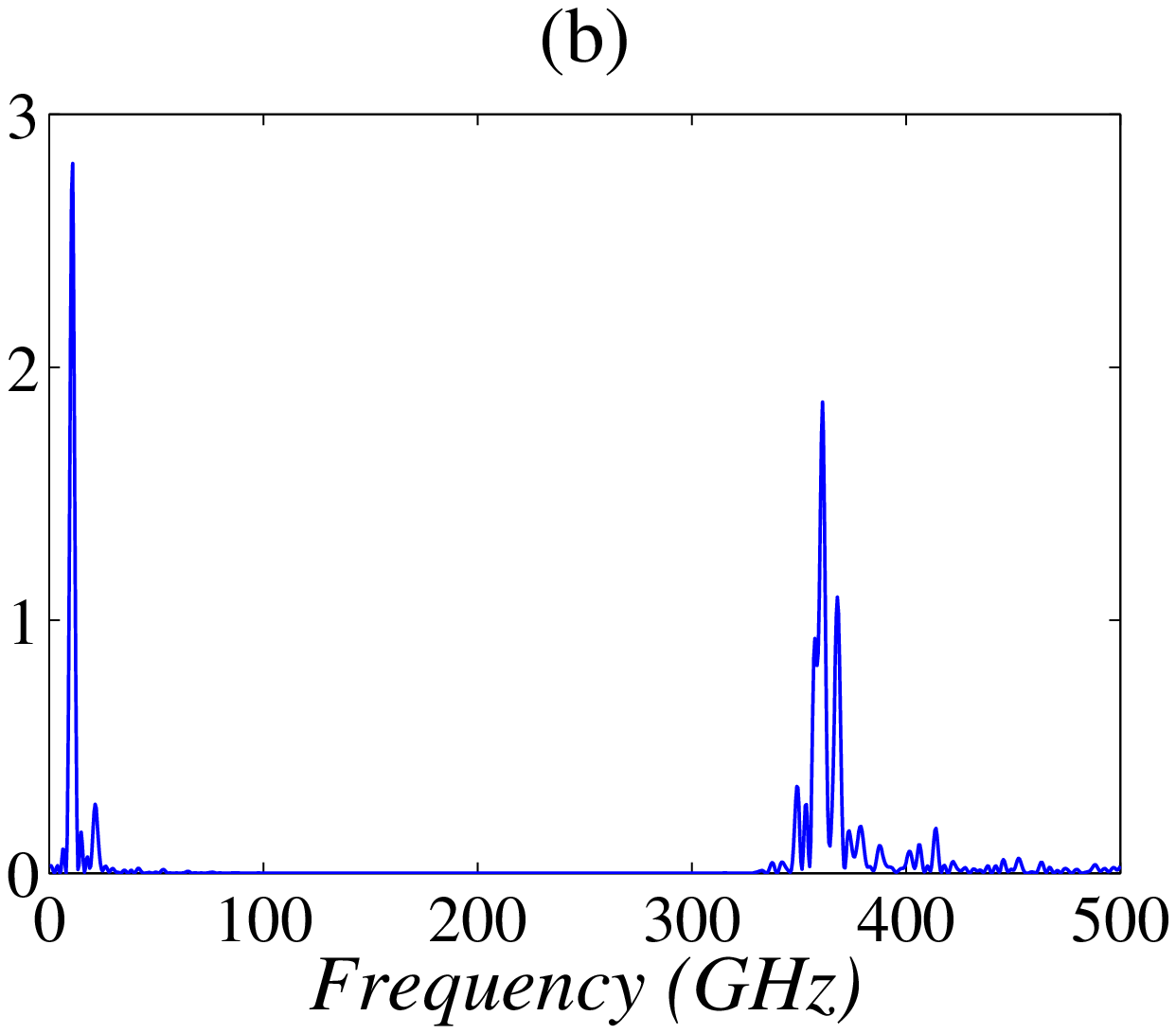}
\vspace{0.2cm} \caption{(a) Total current density vs time during coexisting Bloch and Gunn type oscillations at 70 K. (b) Fourier transform
of the total current density showing two peaks corresponding to coexisting Bloch (0.36 THz) and Gunn type (11 GHz) oscillations. }
\label{fig2}
\end{center}
\end{figure}

 For a voltage bias $\phi=0.05$ ($V = 0.166$ V) and after a short transient that depends on the initial conditions, we observe coexisting BOs of frequency 0.36  THz and Gunn type oscillations of frequency 11 GHz. BOs are stable because $(\gamma_e+\gamma_j)/2<\gamma_{\rm crit}$ and Gunn type oscillations are a consequence of the periodic recycling and motion of electric field pulses from the cathode to the anode. Figure \ref{fig1} shows several snapshots of the field and $|A|$ profiles of the Gunn type oscillation, and Fig. \ref{fig1.1} exhibits the corresponding current density profiles for $\theta=0$. While the amplitude of Gunn-type current oscillation is about $0.03$ in nondimensional units (as seen in Fig. \ref{fig1}(a) for the total current density averaged over the BOs), the BO part of the current oscillation has a larger amplitude of about $0.5$ (as shown in Fig. \ref{fig1}(c) for the modulus of $A$). Figure \ref{fig2} illustrates the total current density (\ref{eqJtot}) of the coexisting $0.36$ THz Bloch and $11$ GHz Gunn type oscillations, respectively. For each lattice temperature, there is a critical curve in the plane of restitution coefficients such that, for $(\gamma_e+\gamma_j)/2>\gamma_{\rm crit}$, BOs disappear after a relaxation time but they persist for smaller values of $(\gamma_e+\gamma_j)$ \cite{BAC11}.

\begin{figure}
\begin{center}
\includegraphics[width=8cm,angle=0]{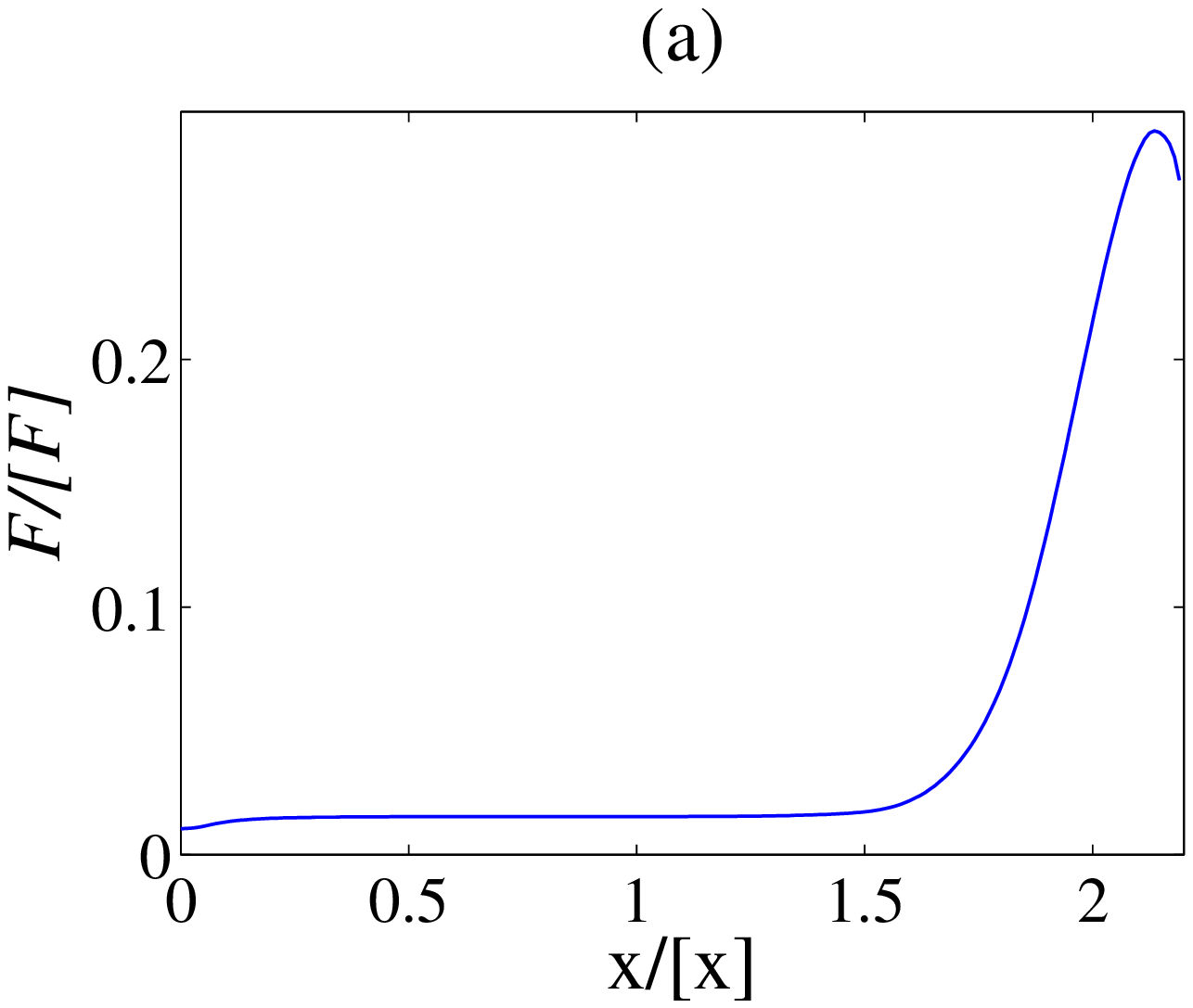}
\includegraphics[width=8cm,angle=0]{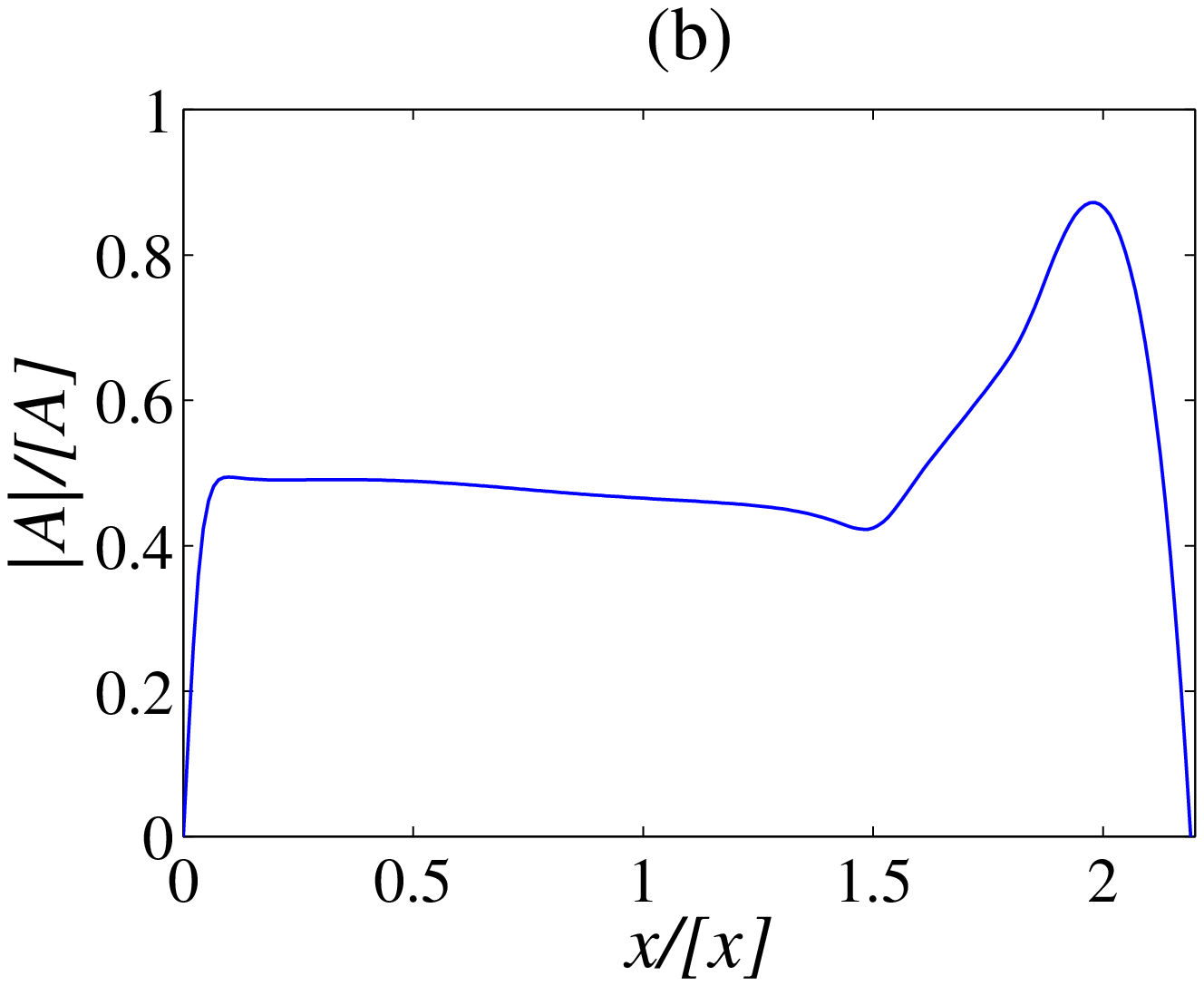}
\vspace{0.2cm} \caption{(a) Electric field profile vs space for the stationary state at 300K. (b) Same for the modulus of the BO complex amplitude. To transform the magnitudes in this figure to dimensional units, use Table \ref{t1}. }
\label{fig3}
\end{center}
\end{figure}

\begin{figure}
\begin{center}
\includegraphics[width=8cm,angle=0]{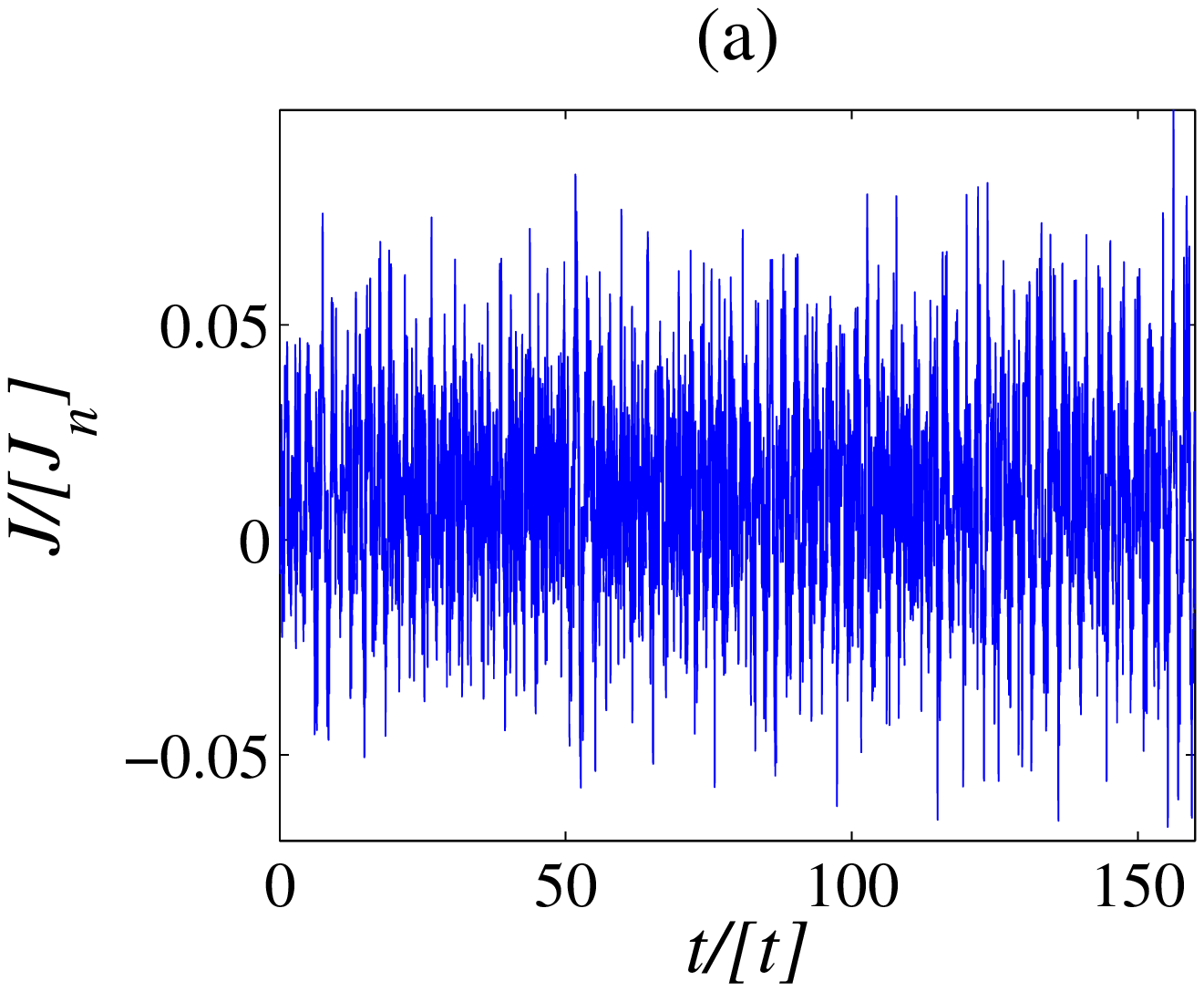}
\includegraphics[width=8cm,angle=0]{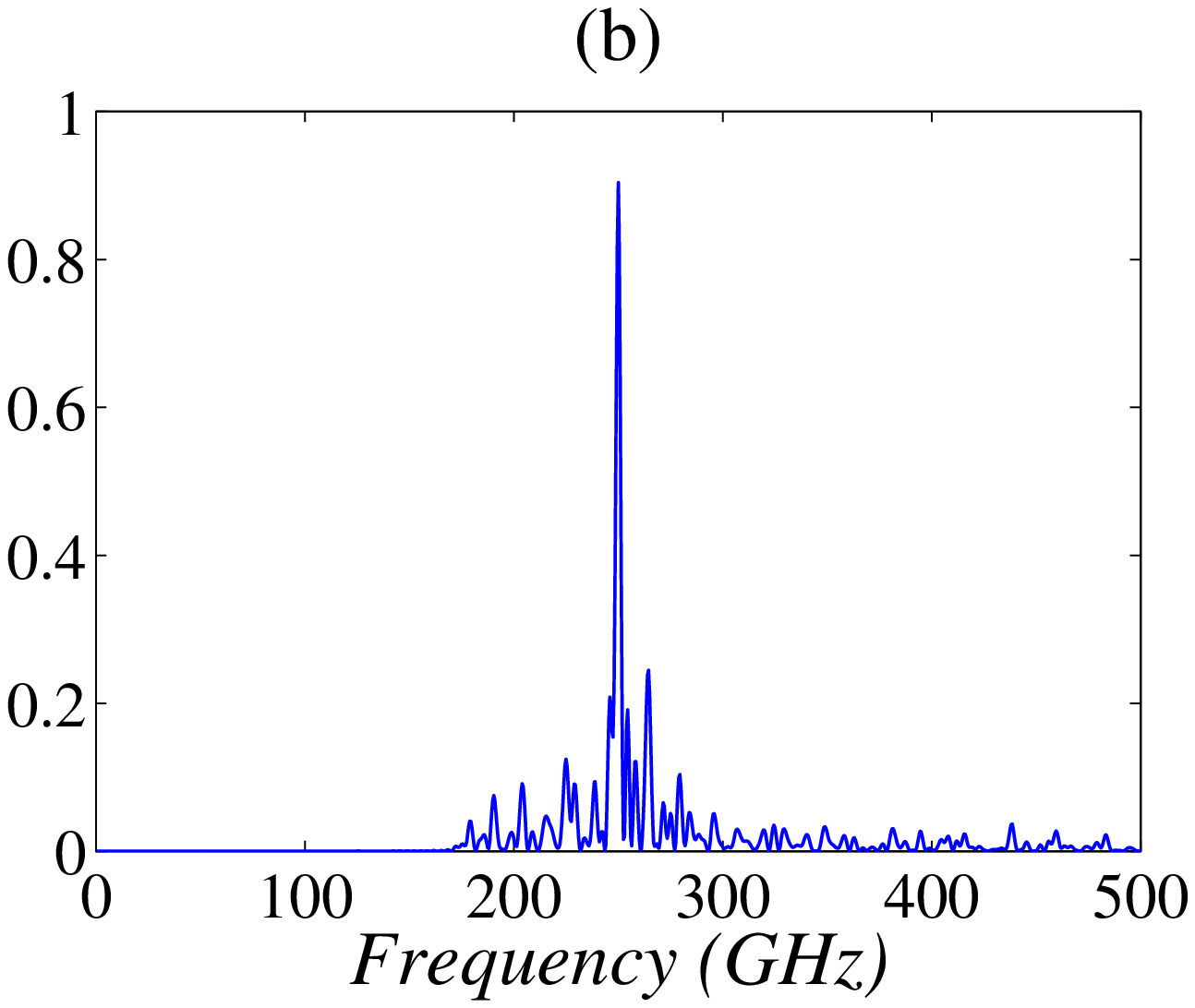}
\vspace{0.2cm} \caption{(a) Total current density vs time during Bloch oscillations at 300K. (b) Fourier transform of the total
current density showing only one peak corresponding to  BOs (0.26 THz). The zero-frequency constant corresponding to the time
average of the total current density has been subtracted.} \label{fig4}
\end{center}
\end{figure}

Figure \ref{fig3} shows the profiles of $F$ and $|A|$ and Fig. \ref{fig4} depicts the total current density at temperature 300 K, with $E_0=0.1529$, for the same values of $\gamma_{e,j}$ and the other parameters. We find BOs but not the slower Gunn type oscillations. Whether Bloch and Gunn type oscillations coexist depends on the relative size of the diffusion (\ref{coefB}) and convection (\ref{coefA}) terms in (\ref{E.dF/dt}) which, in turn, are controlled by the lattice temperature according to (\ref{E0}). There is a critical temperature below which diffusion terms are sufficiently small compared to convective terms in (\ref{ddeF}) and the electric field pulses are then periodically recycled when they reach the anode, originating the Gunn type oscillations. For larger temperatures, Bloch and Gunn type oscillations cannot occur simultaneously: When the electric field pulse reaches the anode, it remains stuck there and the electric field profile becomes the stationary state shown in Fig. \ref{fig3}. Note that the largest peak in the current spectrum in Fig. \ref{fig4}(b) occurs at a lower frequency (0.26 THz) than in the case of lower lattice temperature shown in Fig. \ref{fig2}(b).

\subsection{Spurious convection}
\begin{figure}
\begin{center}
\includegraphics[width=8cm,angle=0]{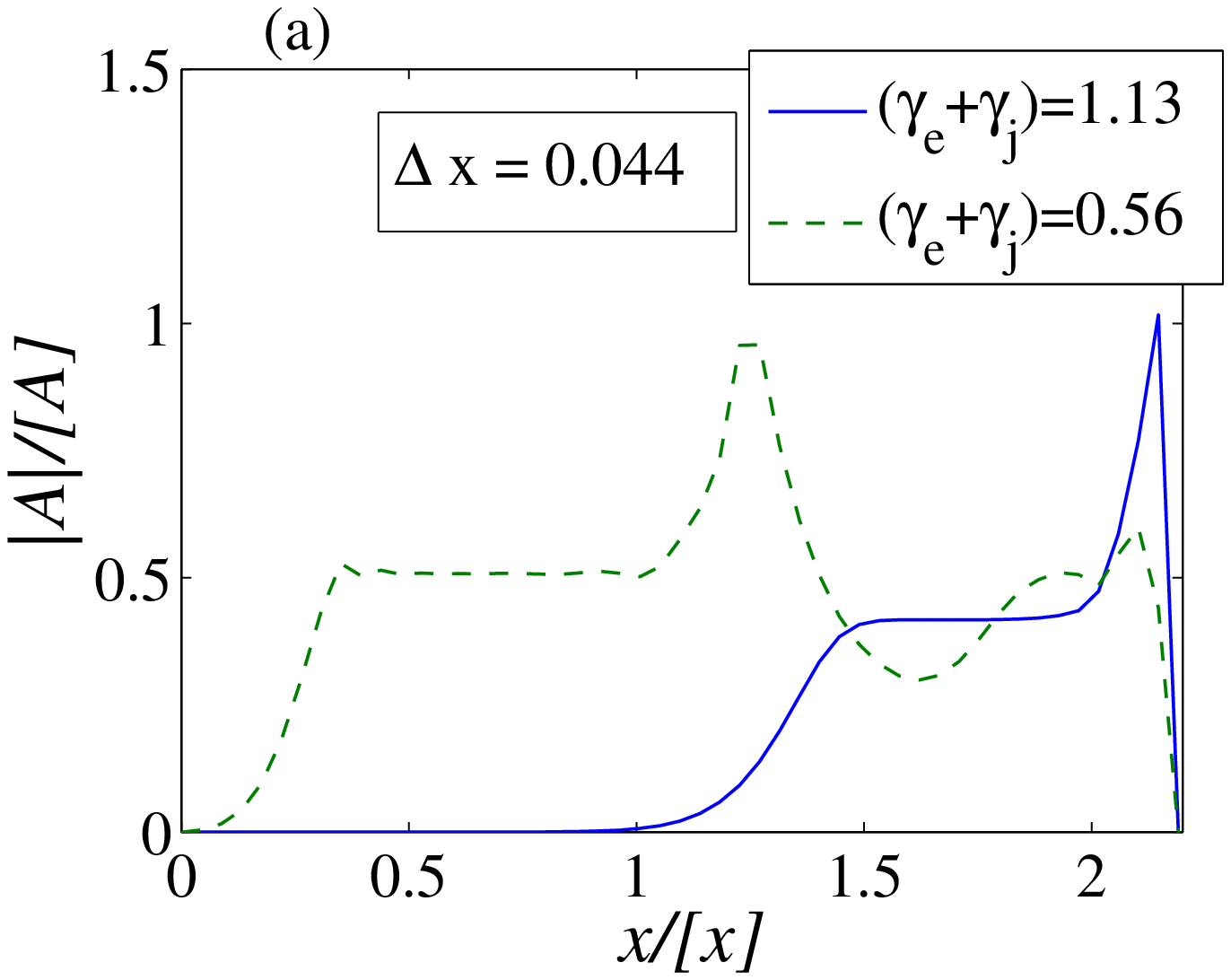}
\includegraphics[width=8cm,angle=0]{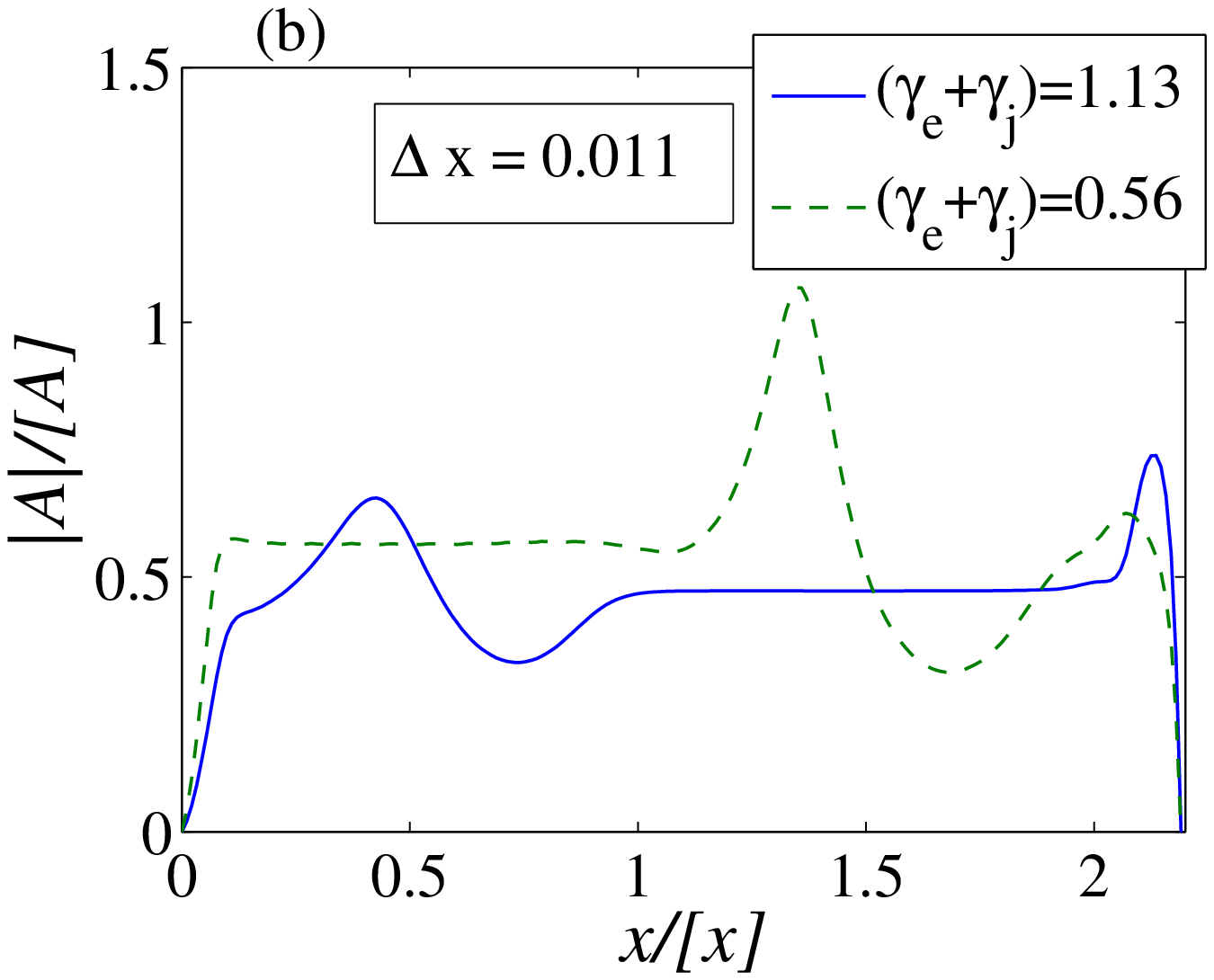}
\vspace{0.2cm} \caption{Modulus of the nondimensional complex BO
amplitude at a given time,  calculated using first-order backward
differences to solve (\ref{eqA}) with different space steps.}
\label{fig_spurious}
\end{center}
\end{figure}

Had we used first-order backward differences for the space derivative in (\ref{eqA}) instead of second-order ones (\ref{2diff}), the computing time to get comparable results would have had to increase significantly: unless we use four times smaller values of $\Delta x$, the scheme with first-order differences gives rise to spurious convection terms in $A$. Fig. \ref{fig_spurious} shows the spatial profile of $|A|$ at a given time during BOs, calculated using first-order backward differences in (\ref{eqA}) for two values of $(\gamma_e+\gamma_j)<(\gamma_e+\gamma_j)_{crit}$. The dashed lines in Figs. \ref{fig_spurious}(a) and (b) show that BOs extend to the whole SL for sufficiently small values of $(\gamma_e+\gamma_j)$, no matter the step size. However, when $(\gamma_e+\gamma_j)$ approaches the critical value from below (solid line), the BOs are confined to part of the SL as in Fig. \ref{fig_spurious}(a) unless the step size is sufficiently small as in Fig. \ref{fig_spurious}(b). Fig. \ref{fig_spurious}(a) shows that, for larger step size, there appears a spurious convection in $A$ that extends the region where $A=0$ from the cathode and it confines the BOs to the region closer to the receiving contact (anode). The spatial interval where $A=0$ moves from cathode towards the anode as time increases. The phenomenon of the spurious convection occurs for restitution coefficients close to their critical values as in Ref. \cite{BACeps11}. 

\section{Convergence of the method} \label{sec:4}
\begin{figure}
\begin{center}
\includegraphics[width=8cm,angle=0]{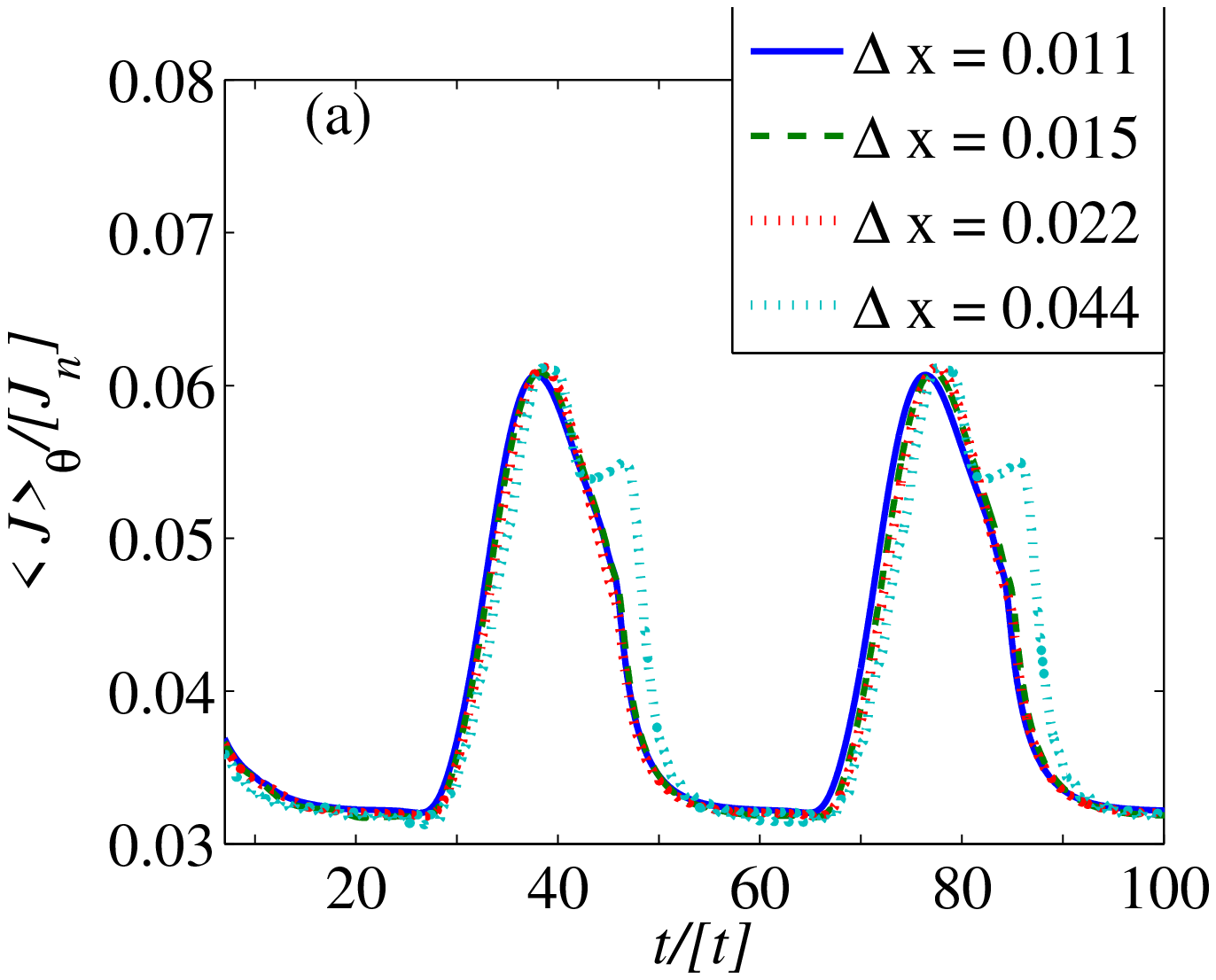}
\includegraphics[width=8cm,angle=0]{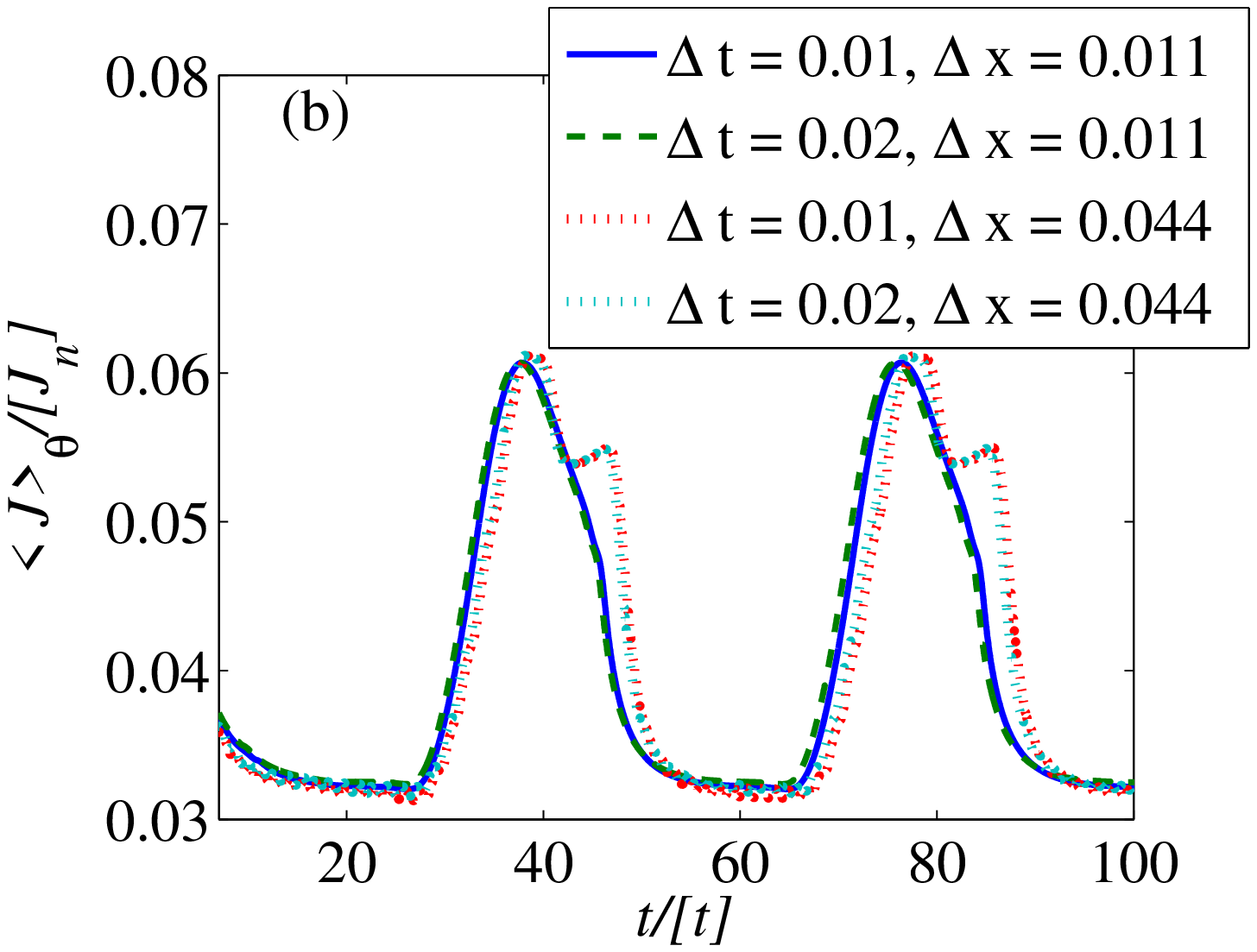}
\vspace{0.2cm} \caption{$\theta$-averaged current density versus
time for different  values of space and time steps. In (a) we use
$\Delta t=0.01$.} \label{fig5}
\end{center}
\end{figure}

We have verified the convergence of the method in terms of the
spatial and temporal mesh size $\Delta x$ and $\Delta t$, by means
of numerical simulations. Figure \ref{fig5} shows the evolution of
$\left<J \right>_\theta$ for different values of $\Delta x$ and
$\Delta t$. Convergence to the solution is more sensitive to the
size of the space step $\Delta x$ than to the size of the time
step $\Delta t$. We observe in Fig. \ref{fig5}(a) that if we use
two or more integration nodes per SL period (i.e., $\Delta x \leq
0.022$), the difference between the numerical solutions is small
and the artifact that appears for large space step, $\Delta
x=0.044$ is eliminated for smaller step sizes. On the other hand,
Fig. \ref{fig5}(b) shows that decreasing the time step by half has
little effect in improving the convergence of the scheme: the
artifact for large $\Delta x=0.044$ still appears whereas for
smaller space steps, the improvement provided by a smaller time
step is quite small.
\begin{figure}
\begin{center}
\includegraphics[width=8cm,angle=0]{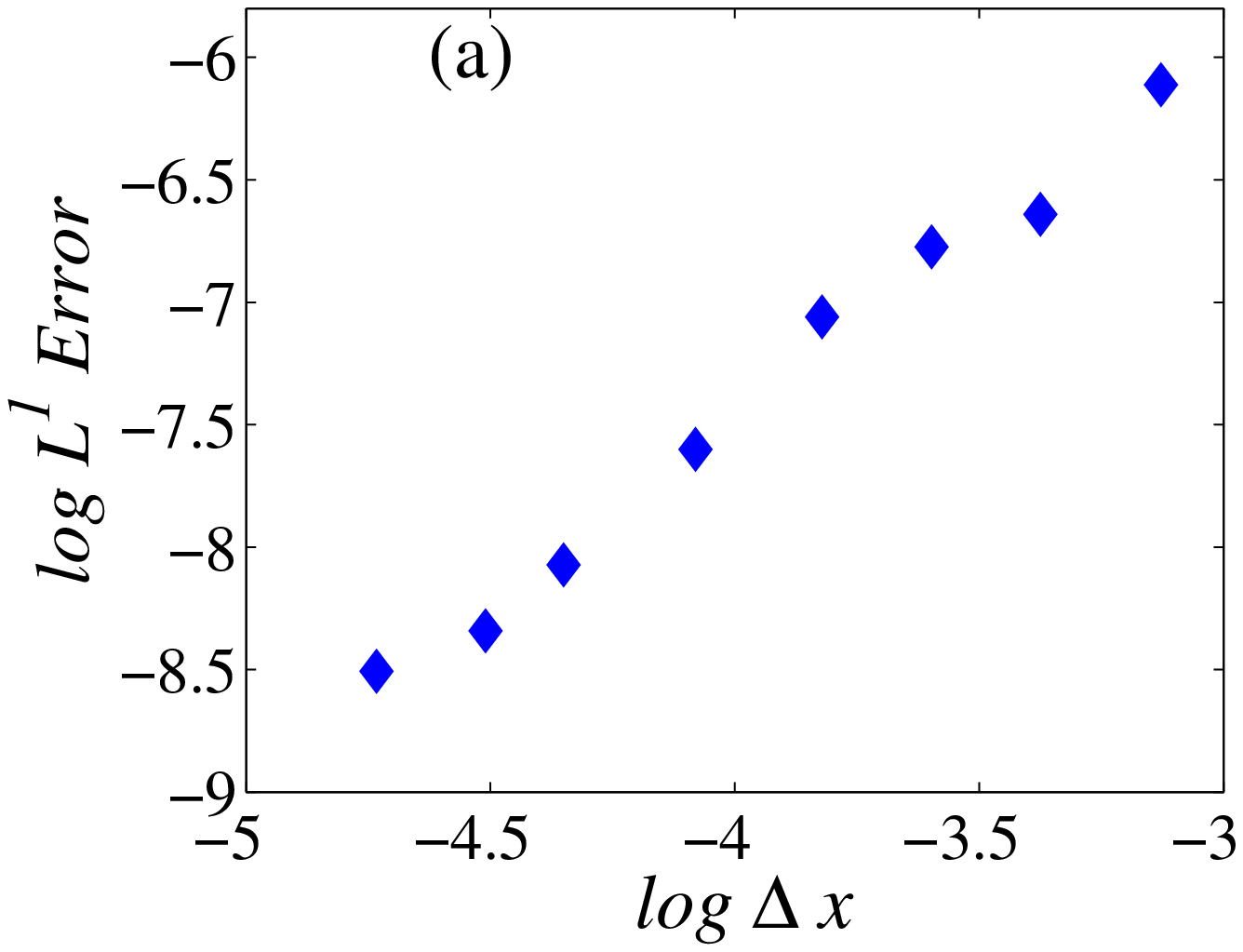}
\includegraphics[width=8cm,angle=0]{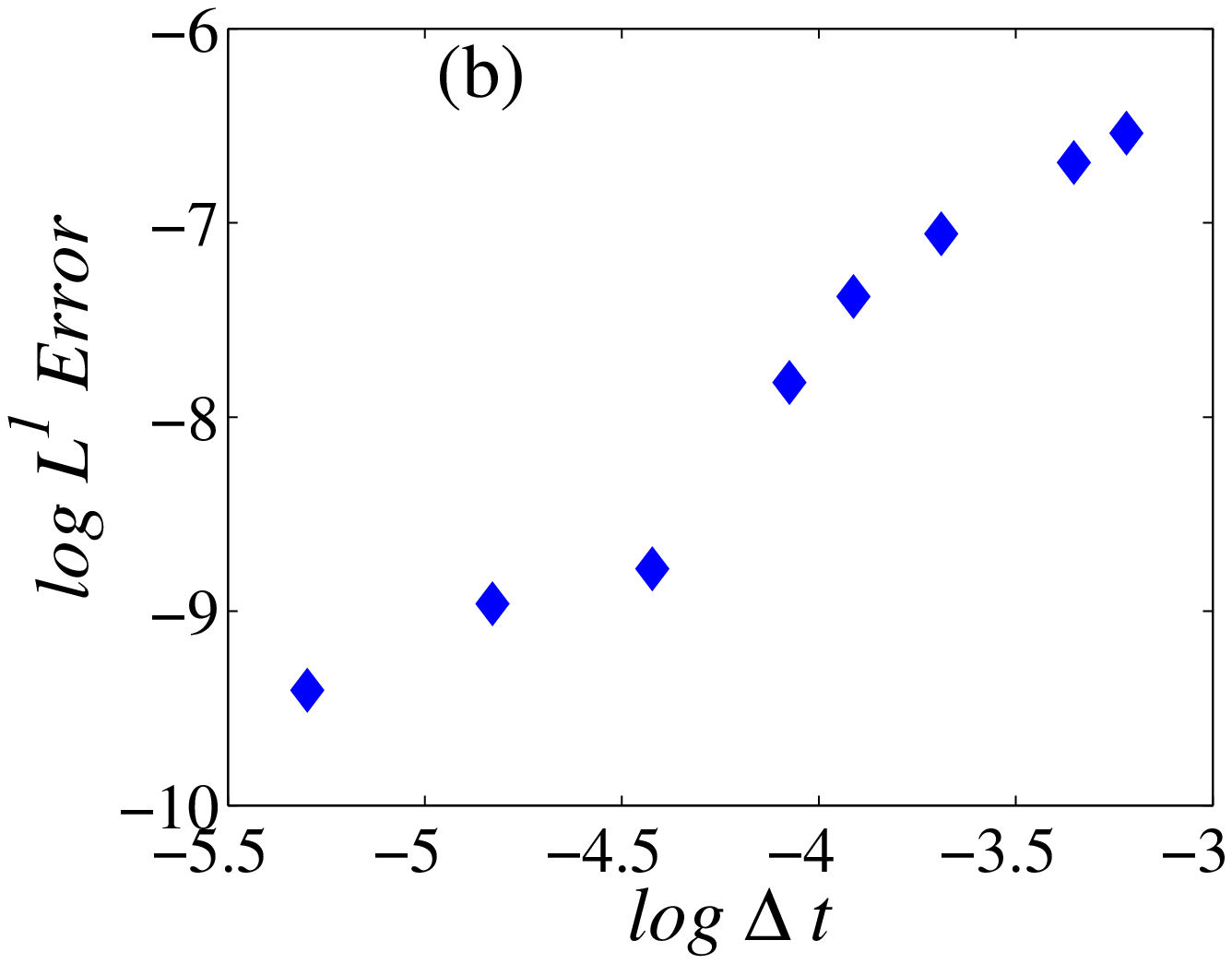}
\vspace{0.2cm} \caption{Error  of $\theta$-averaged current
density (in the $L^1$-norm) for different  values of $\Delta x$
and $\Delta t$. In (a) we use $\Delta t=0.01$. In (b) we use
$\Delta x=0.011$.} \label{fig6}
\end{center}
\end{figure}

We also analyzed the average number of fixed point iterations necessary   for  convergence of the numerical scheme at each time step and checked that the fixed point iteration is contractive. In the case where the solution consists of Gunn-like self-oscillations, only two or three iterations are needed for each time step during most of one period. However when the field pulses reach the SL end and new pulses are created at the cathode, more than three iterations are needed to attain a reliable result. We have observed that the number of iterations depends directly on the size of the time step $\Delta t$. Table \ref{table2} summarizes the main results of the convergence analysis for a simulation of $t=200[t]=376$ ps. It can be seen that for larger $\Delta t$, more fixed point iterations are needed at each time step, therefore the total computation time, for a given $\Delta x$, does not depend too much on $\Delta t$ as long as it is small enough for the convergence of the iterative scheme. 

Figure \ref{fig6} demonstrates the convergence of our numerical scheme. Figures \ref{fig6} (a) and (b) depict the time-averaged error of $\langle J\rangle_\theta$ (in the $L^1$-norm) as a function of the spatial and time discretizations, respectively. We observe
that the convergence order  is approximately quadratic ($\approx
1.88$) in space and ($\approx 1.92$) in time.

\vspace{0.5cm}
\begin{table}[ht]
\begin{center}
\begin{tabular}{|c|c|c|c|c|c|}
  \hline
  $N$ & $\Delta x$ & $\Delta t$ & \# iterations/\# steps (t) & Total \# iterations & Mflops/\# iterations    \\
  \hline
  50 & 0.044 & 0.03 & 3.011 & 20073 & 0.554 $N$  \\
  50 & 0.044 & 0.02 & 2.123 & 21230 & 0.554 $N$  \\
  100 & 0.022 & 0.03 & 2.987 &  19913 & 0.570 $N$  \\
  100 & 0.022 & 0.02 & 2.138 & 21380 & 0.570 $N$  \\
  150 & 0.015 & 0.03 & 2.945 & 19633 & 0.572 $N$  \\
  150 & 0.015 & 0.02 & 2.153 & 21530 & 0.572 $N$  \\
  200 & 0.011 & 0.03 & 2.913 & 19420 & 0.575 $N$   \\
  200 & 0.011 & 0.02 & 2.334  & 23340 & 0.575 $N$   \\
  \hline
\end{tabular}
\caption{Average number of fixed point iterations  and Mflops. } \label{table2}
\end{center}
\end{table}
We can also observe in Table \ref{table2} that the average number of floating point operations is almost proportional to the number of
space integration nodes ($N$). Considering the average number of  iterations per time step, the total computational cost for a  simulation
with $N=200$, $\Delta t=0.02$ and $N_{steps}=10000$ (Cost $\approx 0.575\times 10^6\,  N N_{iterations} N_{steps}$) is of the order of $2.4
\times 10^{12}$ flops. The computation time
in a computer with  Intel(R) Core(TM)2 Duo CPU E6750 @ 2.66 GHz processor is about 30 hours. These estimates show that although the problem is one dimensional, it is rather computationally intensive and therefore it is important to optimize
the numerical algorithm.

\section{Conclusions} \label{sec:5}
We have presented a finite difference method to numerically solve a new type of hydrodynamic equations that arise in the theory of spatially inhomogeneous Bloch oscillations in semiconductor SLs. The hydrodynamic equations describe the evolution of the electric field, electron density and the complex envelope of the Bloch oscillations for the electron current density and the mean energy density. They contain averages over the Bloch phase which are integrals of the unknown electric field. These equations are derived by singular perturbation methods from a Boltzmann-Poisson transport model of miniband SLs with inelastic collisions. Among the solutions of the hydrodynamic equations at low lattice temperature, there are spatially inhomogeneous Bloch oscillations coexisting with moving electric field domains and Gunn-type oscillations of the current. The latter oscillations disappear at higher temperatures (300 K). These novel Bloch-oscillation solutions are found for restitution coefficients in a narrow interval below their critical values and disappear for larger values of the restitution coefficients.

To solve the averaged hydrodynamic equation for the time evolution of the complex BO amplitude, we use an efficient implicit second-order
numerical scheme that uses a fixed-point iteration process to avoid numerical instabilities. In the case of the drift-diffusion equation
for the electric field, only one tridiagonal matrix needs to be inverted to solve the implicit scheme, which results in a greatly decreased
computational time.  Double integrals entering the averaged hydrodynamic equations are calculated by means of expansions in modified Bessel
functions. We use numerical simulations to ascertain the convergence of the method. If the complex amplitude equation is solved using a
first order scheme for restitution coefficients near their critical values, a spurious convection arises that annihilates the complex
amplitude in the part of the superlattice that is closer to the cathode. This numerical artifact disappears if the space step is
appropriately reduced or we use the second-order numerical scheme.\\

{\bf Acknowledgements:} This work has been supported by the MICINN grant FIS2011-28838-C02-01. We thank Conrad Perez (U.\ Barcelona) for
suggesting the use of expansions in series of Bessel functions.

\appendix
\section{Numerical method for obtaining $u$ and $\beta$ }\label{app}
Solving  (\ref{EqNewton1}) by means of the Newton-Raphson method with the following
Jacobian matrix:
\begin{eqnarray}
 \mathcal J  = \left(
    \begin{array}{cc}
      {\partial \hspace{3mm} \over \partial u^{(0)} }\!\left({K_{c\beta} \over K_c} \right) &  {\partial \hspace{3mm} \over \partial \beta^{(0)} }\!\left({K_{c\beta} \over K_c} \right)\! \\
      {\partial \hspace{3mm} \over \partial u^{(0)} }\!\left({K_{s} \over K_c} \right) &  {\partial \hspace{3mm} \over \partial \beta^{(0)} }\left({K_{s} \over K_c} \right) \\
    \end{array}
  \right)\!, \label{Jacobian}
\end{eqnarray}
we obtain the numerical values of $u^{(0)}$ and $\beta^{(0)}$. Thus we need to calculate numerically the integrals (\ref{Int1})-(\ref{Int3}), as well as the following ones:
\begin{eqnarray}
&&K_{c\beta\beta} = {\partial^2 K_c \over \partial (\beta^{(0)})^2 } = \int_{0}^\pi e^{\beta^{(0)}\cos k}\cosh(u^{(0)}k)\cos^2k\, dk, \label{Int4}\\
&&K_{cu\beta} = {\partial^2 K_c \over \partial u^{(0)} \partial \beta^{(0)} } = \int_{0}^\pi e^{\beta^{(0)}\cos k}k \sinh(u^{(0)}k)\,\cos k\, dk,  \label{Int5}\\
&&K_{s\beta} =   {\partial K_s \over \partial \beta^{(0)}} = \int_{0}^\pi e^{\beta^{(0)}\cos k}\sinh(u^{(0)}k)\sin k\cos k\, dk, \label{Int6}\\
&&K_{su} = {\partial K_s \over \partial u^{(0)} } = \int_{0}^\pi e^{\beta^{(0)}\cos k}k\cosh(u^{(0)}k)\sin k\, dk, \label{Int7}\\
&&K_{cu} = {\partial K_c \over \partial u^{(0)} } = \int_{0}^\pi e^{\beta^{(0)}\cos k}k\sinh(u^{(0)}k)\, dk. \label{Int8}
\end{eqnarray}
We also need the integrals (\ref{Int1})-(\ref{Int3}) and (\ref{Int4})-(\ref{Int8}) to obtain $u^{(1)}$ and $\beta^{(1)}$ from (\ref{equ1}) and (\ref{eqbeta1}), and therefore an efficient  method is required for their numerical computation. If we expand $e^{\beta^{(0)} \cos k}$ as a series of modified Bessel functions, these integrals can be expressed as the following series (we omit the argument $\beta^{(0)}$ of the Bessel functions):
\begin{eqnarray}
&&K_c = {\sinh(\pi u^{(0)}) \over u^{(0)}}\left(I_0 + 2\!\left(u^{(0)}\right)^2\sum_{j=1}^\infty (-1)^j {I_j \over j^2 +(u^{(0)})^2} \right)\!, \nonumber\\
&&K_s = {\sinh(\pi u^{(0)}) }\!\left[ {I_0 \over 1+(u^{(0)})^2 } + \sum_{j=1}^\infty (-1)^j I_j\!\left( {1+j \over (j+1)^2 +
(u^{(0)})^2} + {1-j \over (1-j)^2 + (u^{(0)})^2}\right)\!\right]\!,\nonumber\\
&&K_{c\beta} = {\sinh(\pi u^{(0)}) \over u^{(0)}}\!\left[I_1 + 2\left(u^{(0)}\right)^2\sum_{j=1}^\infty  {(-1)^j \over j^2 +
(u^{(0)})^2}\!\left(I_{j+1}+ {jI_j \over \beta^{(0)} }  \right)\!  \right]\!,\nonumber\\
&&K_{c\beta\beta} = {\sinh(\pi u^{(0)}) \over u^{(0)}}\!\left[I_2 + {I_1 \over \beta^{(0)} }+ 2\!\left(u^{(0)}\right)^2\sum_{j=1}^\infty
{(-1)^j \over
j^2 + (u^{(0)})^2}\left(I_{j+2} \right. \right.\nonumber \\
&&\quad\quad \left.\left. + {(2j+1)I_{j+1} \over \beta^{(0)} } + {j(j-1)I_{j} \over (\beta^{(0)})^2 }  \right)\!    \right]\!,\nonumber\\
&&K_{cu\beta} = {\sinh(\pi u^{(0)}) \over u^{(0)}} \sum_{j=1}^\infty (-1)^j{4j^2u^{(0)}  \over (j^2 + (u^{(0)})^2)^2}\!\left(I_{j+1}+ {jI_{j} \over \beta^{(0)} }  \right)\nonumber  \\
&&\quad\quad + {\pi u^{(0)} \cosh(\pi u^{(0)})-\sinh(\pi u^{(0)}) \over (u^{(0)})^2}\!\left[I_1 + 2(u^{(0)})^2 \sum_{j=1}^\infty
{(-1)^j \over j^2 + (u^{(0)})^2} \!\left(I_{j+1}+ {jI_{j} \over \beta^{(0)} } \right)\!\right]\!,\nonumber\\
&&K_{s\beta} = {\sinh(\pi u^{(0)}) }\!\left[ {I_1 \over 1+(u^{(0)})^2 } \right. \nonumber \\
&&\quad\quad \left. + \sum_{j=1}^\infty (-1)^j\!\left( I_{j+1} + {jI_{j} \over
\beta^{(0)}}\right)\! \left({1+j \over (j+1)^2 + (u^{(0)})^2} + {1-j \over (1-j)^2 + (u^{(0)})^2}\right)\! \right]\!,\nonumber\\
&&K_{su} = I_0{\pi(1+(u^{(0)})^2)\cosh(\pi u^{(0)}) - 2u^{(0)}\sinh(\pi u^{(0)}) \over (1+(u^{(0)})^2)^2 } \nonumber \\
&&\quad\quad  + [\pi\cosh(\pi u^{(0)}) - 2u^{(0)}\sinh(\pi u^{(0)}) ]\sum_{j=1}^\infty (-1)^j\!\left( {1+j \over (j+1)^2 + (u^{(0)})^2} \right. \nonumber\\
&&\quad\quad  \left. + {1-j \over (1-j)^2 + (u^{(0)})^2}\right)\!  I_j ,\nonumber\\
&&K_{cu} = {\pi u^{(0)}\cosh(\pi u^{(0)}) - \sinh(\pi u^{(0)}) \over (u^{(0)})^2 }\!\left(I_0 + 2(u^{(0)})^2\sum_{j=1}^\infty
(-1)^j {I_j \over j^2 + (u^{(0)})^2} \right) \nonumber \\
&&\quad\quad     + {\sinh(\pi u^{(0)}) \over u^{(0)}} \sum_{j=1}^\infty (-1)^j {4 j^2 I_j u^{(0)}  \over (j^2 + (u^{(0)})^2)^2}, \nonumber
\end{eqnarray}
where $I_j=I_j(\beta^{(0)})$ is the modified Bessel function of the first kind with index $j$ and argument $\beta^{(0)}$.  Since $|u^{(0)}|$ and $|\beta^{(0)}|$ are less than $3.5$, with only 13 terms of the previous Bessel series we can have an error less than $10^{-4}$ percent.

The numerical value of $u^{(1)}$ and $\beta^{(1)}$ can be obtained from (\ref{equ1beta1}) considering that $f^{B(1)}$ is
\begin{eqnarray}
&&f^{B(1)} = \pi n\left(G_1\,\beta^{(1)} + G_2\,u^{(1)} \right)\!, \label{fB1}
\end{eqnarray}
in which $G_1$ and $G_2$ are:
\begin{eqnarray}
&&G_1(u^{(0)},\beta^{(0)};k) = e^{\beta^{(0)}\cos k + u^{(0)}k }\!\left( {\cos k  \over K_c} - {K_{c\beta}  \over
K_c^2 } \right)\!, \\
&&G_2(u^{(0)},\beta^{(0)};k) = e^{\beta^{(0)}\cos k + u^{(0)}k }\!\left( {k  \over K_c} - {K_{cu} \over K_c^2 } \right)\!.
\end{eqnarray}
From (\ref{fB1}) we can derive $f^{B(1)}_1$ and then $u^{(1)}$ and $\beta^{(1)}$ can be explicitly obtained by solving a linear system of
equations:
\begin{eqnarray}
&&u^{(1)} = {K_c^2 \over n} \frac{\left(Q_{11}\, R_2 - Q_{21}\, R_1 \right)}  {\left(Q_{11}\,Q_{22}-Q_{12}\,Q_{21}\right)}, \label{equ1}\\
&&\beta^{(1)} = {K_c^2 \over n}\frac{\left(Q_{22}\, R_1 -  Q_{12}\, R_2\right) }{\left(Q_{11}\,Q_{22}-Q_{12}\,Q_{21}\right)},
\label{eqbeta1}
\end{eqnarray}
where $R_1$ and $R_2$ are the real and imaginary parts of the RHS of (\ref{equ1beta1}) and the $Q_{ij}$ are:
\begin{eqnarray}
&&Q_{11} = K_c\, K_{c\beta\beta} - \left( K_{c\beta} \right)^2, \nonumber\\
&&Q_{12} = K_c\, K_{cu\beta} - K_{cu}\, K_{c\beta}, \nonumber\\
&&Q_{21} = -K_c\, K_{s\beta}   + K_s\,K_{c\beta}, \nonumber\\
&&Q_{22} = -K_c\, K_{su}  + K_s\,K_{cu}. \nonumber
\end{eqnarray}


\end{document}